\documentclass[aps,twocolumn,showpacs,preprintnumbers,prb]{revtex4}

\usepackage{graphicx}
\usepackage{dcolumn}
\usepackage{bm}

\begin{document}


\title{Electron-hole and plasmon excitations in $3d$ transition metals: {\it Ab
initio} calculations and inelastic x-ray scattering measurements}
\author{I.~G.~Gurtubay$^{1}$, J.~M.~Pitarke$^{1,2}$, Wei Ku$^3$, 
A.~G.~Eguiluz$^{4,5}$, B.~C.~Larson$^5$, J.~Tischler$^5$, 
P.~Zschack$^6$, and K. D. Finkelstein$^7$}
\affiliation{
$^1$Materia Kondentsatuaren Fisika Saila, Zientzi Fakultatea,
Euskal Herriko Unibertsitatea,\\
644 Posta kutxatila, E-48080 Bilbo, Basque Country, Spain\\
$^2$Donostia International Physics Center (DIPC) and Unidad de  F\'\i sica
Materiales
CSIC-UPV/EHU,\\ 
Manuel de Lardizabal Pasealekua, E-20018 Donostia, Basque Country,
Spain\\
$^3$Department of Physics, Brookhaven National Laboratory,
Bldg 510, Upton, NY 11973-5000 and \\
Department of Physics and Astronomy, SUNY Stony Brook, Stony Brook, New
York 11794-3800\\
$^4$Department of Physics and Astronomy, The University of Tennessee,
Knoxville, Tennessee 37996-1200\\
$^5$Condensed Matter Sciences Division, Oak Ridge National
Laboratory, Oak Ridge, Tennessee 37831-6030\\
$^6$Frederick Seitz Materials Research Laboratory, University of
Illinois, Urbana-Champaign, Illinois 61801\\
$^7$CHESS, Cornell University, Ithaca, NY 14853
}
\date{\today}

\begin{abstract}
We report extensive all-electron time-dependent density-functional calculations
and nonresonant inelastic x-ray scattering measurements of the dynamical
structure factor of $3d$ transition metals. For small wave vectors, a plasmon peak is observed which is well described by our calculations. At large wave vectors, both theory and experiment exhibit characteristic low-energy electron-hole excitations of $d$ character which correlate with the presence of $d$ bands below and above the Fermi level. Our calculations, which have been carried out in the random-phase and adiabatic local-density approximations, are found to be in remarkable agreement with the measured dynamical structure factor of Sc and Cr at energies below the semicore onset energy (M-edge) of these materials.
\end{abstract}


\maketitle

\section{introduction}

Electron-hole and plasmon excitations in solids have been investigated for
many years on the basis of the wave-vector and frequency dependent
dynamical electron density response of the solid.\cite{Pnozieres}
The first theoretical investigations of the dynamical density response
function which take into account explicitly the band structure of the solid were
performed during the last decade.\cite{quong93,aryasetiawan94,maddocks94} Since then, first-principles calculations of the
dynamical response of real solids have been reported for simple metals and
semiconductors,\cite{fleszar95,fleszar97,rohlfing00} and also for the noble
metals Cu,\cite{campillo99} Ag,\cite{cazalilla00} and Au,\cite{gurtubay01}
a few transition metals,\cite{schone02,gurtuhcp04}
the transition-metal oxides TiO$_2$\cite{tio2reining,gurtutio2-04} and NiO,
\cite{nio} MgB$_2$,\cite{ku02} and manganites.\cite{grenier05}
Most of these calculations discuss the response of the solid in the 
framework of time-dependent density functional theory (TDDFT) within the 
random-phase approximation (RPA) and  also incorporating the many-body effects,
i.e. exchange and correlation effects, using some local-density approximation (LDA). In general, 
these approximations have proven to 
be quite effective  when studying excitations which do not involve localized
states.

Vast {\it et al.}\cite{tio2reining} have recently reported 
{\it ab initio} calculations of the electron energy-loss spectrum of 
the transition metal oxide TiO$_2$,
focusing on the excitation from the Ti 3p semicore levels. They have compared
their results with electron energy-loss spectroscopy (EELS) measurements and have found that
the inclusion of crystal local-field effects (CLFE), which arise when the microscopic electric field varies rapidly over the unit cell,\cite{adler62} turns out to be crucial
for a faithful description of the experimental spectrum for small wave vectors.

The dynamical electron density response of solids 
 can be probed with the use of optical
absorption experiments,\cite{raether80} EELS,\cite{misell73,raether80} and
inelastic x-ray scattering (IXS) techniques.\cite{platzman73} In particular, the advent of synchrotron sources and
high-resolution IXS facilities have allowed detailed investigations of the
processes responsible for short-range many-body effects     
at large wave vectors.\cite{platzman92,schulke95,larson96} 
Macrander {\it et al.}\cite{macrander96} have reported the first
IXS measurements showing a strong spectral peak for excitations from
 Ti 3$p$ semicore states to empty states above the Fermi level. In particular,
these authors measured the dynamical structure factor for Ti and TiC
single crystals for intermediate and large momentum transfers, and in the 
case of TiC they compared their measurements to local-density-functional calculations with no inclusion of crystal local-field effects. 
Montano {\it et al.}\cite{montano02} included 
these effects in the study of the dynamical structure factor of SiC 
for large wave vectors, 
finding the inclusion of crystal local-field effects
to be very important in order to explain the measured IXS spectrum.
More recently, Gurtubay {\it et al.}\cite{gurtutio2-04} have found that large
crystal local-field effects in the dynamical structure factor of  
TiO$_2$   yield a sharp loss peak in the low-energy regime ($\sim$14~eV), 
which is also present in the experimental IXS spectra.

In this paper, we report extensive all-electron TDDFT calculations and nonresonant IXS measurements of the dynamical
structure factor of various $3d$ transition metals with $d$ states below and
above the Fermi level. For small wave vectors, a plasmon peak is observed which is well described by our calculations. At large wave vectors, both theory and
 experiment exhibit a sharp peak which is absent for small wave vectors and 
which we argue to be originated in the presence of direct $d$-to-$d$ 
transitions involving $d$ states below and above the Fermi level. 
Comparison between theory and experiment indicates that at low energies the 
basic physics of the dynamical response is contained in the
 random-phase approximation (RPA) 
as long as the crystal local-field effects are taken into account. 
Furthermore, our results indicate that the inclusion of many-body effects through an adiabatic LDA (ALDA) kernel yields an unexpectedly good description of the electronic response of valence electrons of predominant $d$ character.

\section{Theory}

\subsection{Single-particle Bloch states: LAPW basis}

The starting point of our calculations is a set of well-converged Bloch states
$\phi_{{\bf k},n}$ and energies $\varepsilon_{{\bf k},n}$, which are the
eigenfunctions and eigenvalues of the Kohn-Sham equation of
density-functional theory (DFT).\cite{KS65,dreizler90} For an accurate description of the dynamics 
of transition metals with narrow $d$ bands, we expand these Bloch
states in a linearized augmented plane wave (LAPW) basis,\cite{singh94,wien97} as follows
\begin{equation}
\label{lapw1}
\phi_{{\bf k},n}({\bf r}) = \sum_{\bf G} C_{{\bf k},n} ({\bf G})
\psi_{{\bf k}+{\bf G}}^{\rm LAPW}({\bf r}),
\end{equation}
${\bf G}$ being vectors of the reciprocal lattice.

The choice of the basis functions $\psi_{{\bf k}+{\bf G}}^{\rm LAPW}({\bf r})$ is
made by partitioning the primitive unit cell into non-overlapping muffin-tin 
spheres centered on the atomic nuclei, where the electronic environment retains  an atomic-like nature, 
and the interstitial region, where electrons may be described by
plane waves. The full effective potential entering the Kohn-Sham equation of
DFT is expressed accordingly, i.e., it is expanded in lattice harmonics inside
the atomic spheres and it is described by using a Fourier series in the
interstitials. The maximum reciprocal-lattice vector ${\bf
G}_{max}$ that we consider in the expansion of Eq.~(\ref{lapw1}) will be dictated
by the cut-off parameter $R_{MT}\times G_{max}$, $R_{MT}$ being the radius
of the smallest muffin-tin sphere in the unit cell.
 
\subsection{Kohn-Sham density-response function}

Once we have an accurate description of the Kohn-Sham single-particle Bloch
states $\phi_{{\bf k},n}$ and energies $\varepsilon_{{\bf k},n}$, we evaluate the
density-response function $\chi^0({\bf r},{\bf r}',\omega)$ of noninteracting
electrons moving in the effective Kohn-Sham potential of DFT.\footnote{The density-response function of a many-electron system yields the electron density induced at a given point of space by an arbitrary (small) frequency-dependent external potential.} Since this
quantity must keep the periodicity of the solid, one can write
\begin{equation}\label{eq8}
\chi^0({\bf r},{\bf r}',\omega)=\frac{1}{\Omega}\sum_{\bf k}^{BZ}\sum_{{\bf
G},{\bf G}'}{\rm e}^{{ i}({\bf k}+{\bf G})\cdot{\bf r}}{\rm
e}^{-{ i}({\bf k}+{\bf G}')\cdot{\bf r}'}\chi_{{\bf G},{\bf G}'}^0({\bf
k},\omega),
\end{equation}
where ${\bf k}$ is a vector in the first Brillouin zone (BZ) and the Fourier
coefficients $\chi_{{\bf G},{\bf G}'}^0({\bf k},\omega)$ are given by the following expression:
\begin{eqnarray}\label{eq9}
\chi_{{\bf G},{\bf G}'}^0({\bf k},\omega)=\frac{1}{ \Omega}\sum_{\bf
k'}^{\rm BZ}\sum_{n,n'} \frac{f_{{\bf k'},n}-f_{{\bf k'}+{\bf k},n'}}
{\varepsilon_{{\bf
k'},n}-\varepsilon_{{\bf k'}+{\bf k},n'} +\hbar(\omega + { i}\eta)}\cr\cr
\times\langle\phi_{{\bf k'},n}|e^{-{ i}({\bf k}+{\bf G})\cdot{\bf
r}}|\phi_{{\bf k'}+{\bf k},n'}\rangle
\langle\phi_{{\bf k'}+{\bf k},n'}|e^{{ i}({\bf k}+{\bf G}')\cdot{\bf
r}}|\phi_{{\bf k'},n}\rangle.
\end{eqnarray}
Here, $\Omega$ represents the normalization volume, $\eta$ is a positive
infinitesimal, and $f_{{\bf k},n}$ are Fermi factors, which at zero temperature
($T=0$) are simply given by the Heaviside step function
\begin{equation}
f_{{\bf k},n}=2\,\Theta(\varepsilon_F-\varepsilon_{{\bf k},n}),
\end{equation}
$\varepsilon_F$ being the Fermi energy of the solid.

The major task in the calculation of the Fourier coefficients $\chi_{{\bf G},{\bf
G}'}^0({\bf k},\omega)$ resides in the computation of the matrix elements entering Eq.~(\ref{eq9}). Using Eq.~(\ref{lapw1}), both the
integration over angular variables and the sum over reciprocal-lattice vectors
can be done analytically, and one finds explicit expressions for the matrix
elements in terms of one-dimensional integrals involving basis radial functions
and spherical Bessel functions.\cite{weiphd}

\subsection{Interacting density-response function}

In the framework of TDDFT,\cite{runge-gross84,gross-kohn85,petersilka96} the {\it
exact} density-response function $\chi({\bf r},{\bf r}',\omega)$ of a periodic solid of electron density $n_0({\bf
r})$ can be expanded in a Fourier series of the form of Eq.~(\ref{eq8}),
with the Fourier coefficients $\chi_{{\bf G},{\bf G}'}({\bf k},\omega)$ given by the
following expression: 
\begin{eqnarray}\label{eq:XGG}
&&\chi_{{\bf G},{\bf
G}'}({\bf k},\omega)=\chi^0_{{\bf G},{\bf G}'}({\bf k},\omega)+
\sum_{{\bf G}_1,{\bf G}_2}\,\chi^0_{{\bf G},{\bf G}_1}({\bf
k},\omega)
\cr\cr
&&\times\left\{v_{{\bf G}_1}({{\bf k}})\delta_{{\bf G}_1,{\bf G}_2}+f_{{\bf
G}_1,{\bf G}_2}^{\rm XC}[n_0]({\bf k},\omega)\right\}\chi_{{\bf
G}_2,{\bf
G}'}({\bf k},\omega),\cr\cr
&&
\end{eqnarray}
where $\chi^0_{{\bf G},{\bf G}'}({\bf k},\omega)$ are the Fourier
coefficients of Eq.~(\ref{eq9}), $v_{\bf G}({{\bf k}})=4\pi e^2/{|{\bf k}+{\bf G}|^2}$ is the
Fourier transform of the bare Coulomb potential, and $f_{{\bf G},{\bf G}'}^{\rm
XC}[n]({\bf k},\omega)$ are the Fourier coefficients of the functional derivative of
the time-dependent exchange-correlation (XC) potential of TDDFT. In the RPA, the XC kernel $f_{{\bf G},{\bf G}'}^{\rm
XC}[n]({\bf k},\omega)$ is taken to be zero; in the ALDA, it is approximated by an adiabatic kernel
of the form
\begin{equation}\label{fxcgg}
f_{{\bf G},{\bf G}'}^{\rm XC}({\bf
k},\omega)=\int d{\bf r}\,{\rm e}^{-i({\bf G}-{\bf G}')\cdot{\bf
r}}\,\left.{dV_{\rm XC}(n)\over
dn}\right|_{n=n_0({\bf r})},
\end{equation}
where $V_{\rm XC}(n)$ is the XC potential of a homogeneous electron
gas of density $n$.

\subsection{Dynamical structure factor and x-ray inelastic scattering cross
section}

The dynamical structure factor $S({\bf q},\omega)$ of a many-electron system,
which determines the electron-density fluctuations at a given wave vector ${\bf
q}$ and frequency $\omega$, is connected to the imaginary part of the
density-response function $\chi({\bf r},{\bf r}',\omega)$ through the fluctuation-dissipation theorem.\cite{callen51}
In the case of a periodic solid at $T=0$, one finds
\begin{equation}
S({\bf k}+{\bf G},\omega)=-2\,\hbar\,\Omega\,{\rm Im}\chi_{{\bf G},{\bf
G}}({\bf
k},\omega),
\end{equation}
${\bf k}$ being a wave vector in the first BZ.

Within the first-Born approximation, the double differential scattering
cross-section for x rays to transfer momentum $\hbar({\bf k}+{\bf G})$ and
energy
$\hbar\omega$ to a periodic solid is given by
\begin{equation}\label{sqw}
{d^2\sigma\over d\Omega d\omega}=\left({e^2\over
m_ec^2}\right)^2\,({\bf
e}_i\cdot{\bf
e}_f)^2\,\left({\omega_f\over\omega_i}\right)\,S({\bf
k}+{\bf G},\omega),
\end{equation}
where (${\bf e}_i$, ${\bf e}_f$) and ($\omega_i$, $\omega_f$) refer to
the polarization vector and frequency of the incident and scattered
photon, respectively.

\subsection{Dielectric matrix and macroscopic dielectric function}

In order to investigate the contribution to the energy-loss spectrum coming from the excitation of plasmons, it is convenient to consider the wave-vector and frequency dependent inverse dielectric matrix of a periodic solid which yields the potential induced at a given point of space by an arbitrary (small) frequency-dependent external potential. One finds
\begin{equation}\label{eps}
\epsilon_{{\bf G},{\bf G}'}^{-1}({\bf k},\omega)=
\delta_{{\bf G},{\bf G}'}+v_{ {
\bf G}}({\bf k})\chi_{{\bf G},{\bf G}'}({\bf k},\omega),
\end{equation}
with ${\bf k}$ being a vector in the first Brillouin zone.

For ${\bf k}$ wave vectors in the first Brillouin zone, the macroscopic dielectric function is defined as follows
\begin{equation}
\epsilon_M({\bf k},\omega)=1/\epsilon_{0,0}^{-1}({\bf k},\omega),
\end{equation}
where $\epsilon_{0,0}^{-1}({\bf k},\omega)$ represents the first diagonal element of the inverse dielectric matrix of Eq.~(\ref{eps}). In the absence of crystal local-field effects (CLFE), the off-diagonal elements of the inverse dielectric matrix are negligible and the macroscopic dielectric function coincides, therefore, with the first diagonal element of the dielectric matrix, $\epsilon_{{\bf G},{\bf G}}({\bf k},\omega)$, which in the RPA is given by the following expression:
\begin{equation}\label{eps0}
\epsilon_{{\bf G},{\bf G}'}^{RPA}({\bf k},\omega)=
\delta_{{\bf G},{\bf G}'}-v_{ {
\bf G}}({\bf k})\chi_{{\bf G},{\bf G}'}^0({\bf k},\omega).
\end{equation}

Collective excitations (plasmons) are known to occur at energies where both the real and the imaginary part of the macroscopic dielectric function are close to zero. On the other hand, the diagonal elements of the imaginary part of the RPA dielectric matrix are known to represent a measure of the number of states available for real transitions involving a given momentum transfer $\hbar{\bf q}$ and energy transfer $\hbar\omega$.  

\section{Experiment}

Nonresonant inelastic x-ray scattering measurements of the dynamical
structure factor  $S({\bf q},\omega)$ of single crystals Sc and Cr have been obtained using the UNICAT undulator
Beam Line on the Advanced Photon Source (APS) at Argonne National
Laboratory. The measurements were made in symmetric 
reflection geometry, with 7.6 keV
x-rays, a range of wave vectors from 0.5 to 3.98 ${\rm\AA}^{-1}$, and an energy
resolution of $1.1\,{\rm eV}$.
The shape-asymmetry of the tails of
the quasi-elastic scattering near zero energy transfer for the analyzer
and scattering geometry used was determined directly by measurements
made on wide band-gap ($>$~10~eV) insulators (e.g. LiF, CaF2), 
which indicated the tails to be
symmetric for losses greater than 2.5 eV.  The
quasi-elastic contribution to the scattering for Sc and Cr was then
removed
using measurements on the energy gain side of the quasi-elastic
peak and restricting the data to energy losses of 2.5 eV or
greater.  The rapid falloff of the quasi-elastic contribution
rendered the tails insignificant for energy losses greater than
6-8 eV for the conditions used.

In order to reduce to absolute units the APS spectra of the transition metals Sc and Cr, we have first performed an absolute measurement on Al.\cite{alnorm} For measurements that do not extend past the Al L-edge at
$\sim 72\,{\rm eV}$, only the three valence electrons of this metal contribute to the
first-moment $f$ sum-rule
\begin{equation}
\label{sqwfsum}
\int_0^{\omega_{max}}{\mathrm d}\omega\,\omega
\;s({\bf q},\omega)=\alpha\,\frac{n_0\,\pi}{m_e}\,{q}^2.
\end{equation}
Here, $n_0$ is the average electron density of the $3s^2$ $3p^1$ valence
electrons in Al, and $s({\bf q},\omega)$ is the reduced dynamical structure factor: $s({\bf q},\omega)=S({\bf q},\omega)/(\hbar\Omega)$. According to
Eq.~(\ref{sqw}), the dynamical structure factor should be proportional to the relative intensity $I({\bf q},\omega)/I_0$ of the scattered beam, $I_0$ being the incident beam power; hence,
\begin{equation}\label{ratio}
s({\bf q},\omega)=C\,I({\bf q},\omega)/I_0.
\end{equation}
The maximum energy $\hbar\omega_{max}$ entering the left-hand side of Eq.~(\ref{sqwfsum}) must be below the 72 eV
semicore onset energy and the coefficient $\alpha$ represents the relative
contribution of valence  $3s^2$ $3p^1$ electrons to the $f$ sum-rule for
energies below $\hbar\omega_{max}$. We have carried out {\it ab initio}
pseudopotential calculations
of the dynamical structure factor of Al, and for $\hbar\omega_{max}=65\,{\rm eV}$
we have found $\alpha=0.95$.

Introducing Eq.~(\ref{ratio}) into Eq.~(\ref{sqwfsum}) one finds the calibration factor $C$, which we have derived from our IXS measurements of the $I({\bf q},\omega)/I_0$ ratio in an Al sample. (See Ref.~\onlinecite{alnorm} for details.)
Since this calibration factor is inversely proportional to the effective volume of material 
sampled by the x-ray beam 
(and is independent of the atomic configuration of the material) and for thick samples and
symmetric Bragg geometry the volume ratio of the material sampled in different targets is given by the ratio of the corresponding linear absorption coefficients, the dynamical structure factor $s_X({\bf q},\omega)$ of an arbitrary ($X$) target can be obtained from the following expression:
\begin{equation}
s_X({\bf q},\omega)=C\,\mu_{X}/{\mu_{Al}}\,I_X({\bf
q},\omega)/I_0,
\end{equation}
with $\mu_{ X}$ and ${\mu_{\mathrm Al}}$ being the linear absorption coefficients of the $X$ material and Al, respectively. 

As suggested
in Ref.~\onlinecite{alnorm}, analogous (albeit sample geometry and q-dependent)
expressions for the ratio of the effective scattering volumes can be
derived for transmission geometry and non-symmetric reflection
scattering geometry.

\section{Results and discussion}

In this section, we first present TDDFT calculations and nonresonant IXS
measurements of the dynamical structure factor of the hexagonal close-packed
(hcp) Sc and the body-centered cubic (bcc) Cr. Then, we present a comparison of the dynamical response of the transition metals Sc, Ti, V, Cr, Fe, Co, and Ni, all belonging to the $3d$ series, and we focus on the evolution of a low-energy peak that is visible at large wave vectors and which is originated in the presence of $d$-to-$d$ transitions involving $d$-states below and above the Fermi level. 

All the calculations presented below have been carried out by first expanding Bloch states in an LAPW basis, as in Eq.~(\ref{lapw1}), with a cutoff parameter of $R_{MT}\times G_{max}=8$, using 
spherical harmonics inside the atomic spheres
up to $l=10$, and including $3s$ and $3p$ states as semicore states. We have then solved
self-consistently the Kohn-Sham equation of DFT in the local-density
approximation (LDA), with use of the Perdew-Wang parametrization\cite{Pwang} of the
Ceperley-Alder XC energy of a uniform electron gas,\cite{CA80} we have calculated the Fourier coefficients $\chi_{{\bf G},{\bf
G}}^0({\bf k},\omega)$ from Eq.~(\ref{eq9}) by keeping unoccupied KS states up to 7.5 Ry, and we have finally solved Eq.~(\ref{eq:XGG}) either in the RPA or the ALDA with a finite damping parameter
$\eta=0.65\,{\rm eV}$.

In the case of the hexagonal close-packed Sc, Ti, and Co, the coefficients $\chi_{{\bf G},{\bf G}}({\bf k},\omega)$ have been
evaluated for wave vectors perpendicular to the hexagonal plane, i.e., along the
(001) direction, and well-converged results have been obtained by using a
$8\times8\times16$ sampling of the BZ corresponding to 90 points in the
irreducible BZ (IBZ). In the case of the body-centered cubic Cr, V, and Fe, calculations have been performed for wave vectors along the (100), (110), and
(111) directions, and well-converged results have been obtained by using a $10\times10\times10$ sampling of the BZ corresponding to 47 points in the IBZ. Finally, for the face-centered cubic Ni the coefficients $\chi_{{\bf G},{\bf G}}({\bf k},\omega)$ have been evaluated for wave vectors along the (100) direction, and well-converged results have been obtained by using a $12\times12\times12$ sampling of the BZ corresponding to 72 points in the IBZ.

In order to investigate the impact of crystal local-field efects, we have first
neglected these effects by considering only the diagonal elements of the
Kohn-Sham density-response matrix entering Eq.~(\ref{eq:XGG}) (diagonal calculation), and
we have then solved the matrix equation with a given number of {\bf G} vectors
(full calculation). Well-converged results have been obtained with the use of 35,
43 and 51 {\bf G} vectors in hcp, bcc and fcc structures, respectively.
\footnote{The actual number of {\bf G} vectors needed for convergence depends on the spectral feature. While the semicore is the feature which needs the largest numbers, fewer {\bf G} vectors are required in the valence region.}

\subsection{Sc}

\begin{figure}[htb]
\includegraphics*[width=0.95\linewidth]{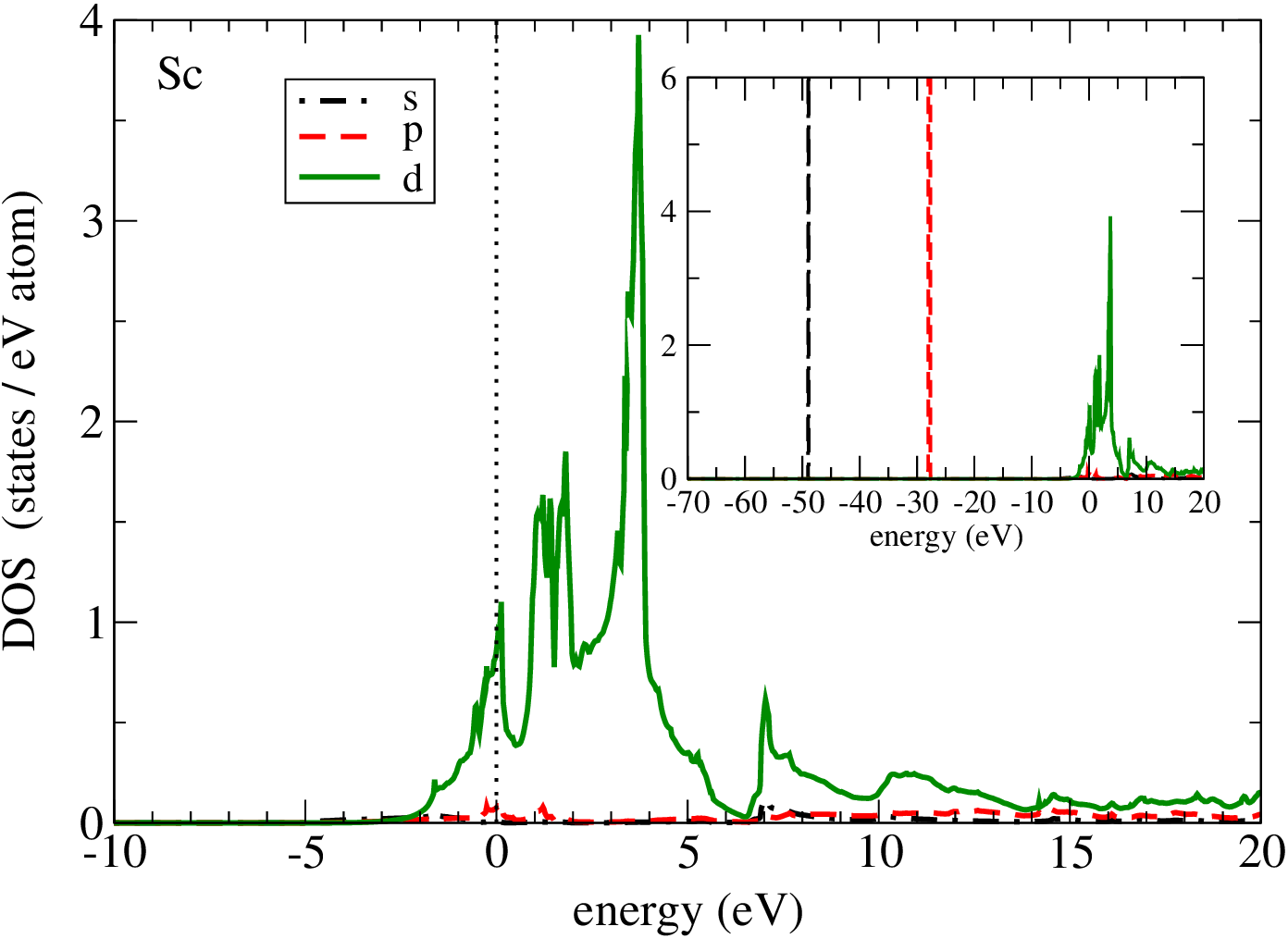} 
\caption{\label{dosscti}
(Color online)
Density of states (DOS) of Sc, at energies near the Fermi level (dotted line) for
each component of the angular momentum inside the atomic sphere.
Dash-dotted black, dashed red and solid green lines denote the number of states per atom and per eV
with $s$, $p$, and $d$ character, respectively.
The inset shows a wider energy range, in which the 3$s$ (dash-dotted black line)
and 3$p$ (dashed red line) semicore states can be distinguished.}
\end{figure}

Sc ([Ar] 3$d^1$ 4$s^2$) being the first of the transition metals has a single
$3d$ electron per atom. Hence, the $3d$ band is located below and above the
Fermi level, as shown by the density of states (DOS) drawn by a solid green curve in Fig.~\ref{dosscti}. The inset of this figure shows the location of
 the 3$s^2$ (dash-dotted black line) and  3$p^6$ (dashed red line) 
semicore  bands, which are
located at  $\sim$49~eV and $\sim$28~eV below de Fermi
level, respectively.

Early measurements of the energy-loss spectra of high-energy electrons in Sc were reported by Brousseau-Lahaye {\it et al.} for small wave
vectors.\cite{sceels75} A plasmon peak was identified at $\sim13\,{\rm eV}$, which
agrees with the expected plasmon energy of valence electrons 
($3d^14s^2$) in Sc.
Pseudopotential-based first-principles calculations of the energy-loss function of Sc were first reported
by Sch\"{o}ne and Ekardt\cite{schone02} at small and intermediate wave vectors and more
recently by Gurtubay {\it et al.}.\cite{gurtuhcp04} A detailed analysis of the plasmon energy
dispersion of this material was reported in Ref.~\onlinecite{gurtuhcp04}, demonstrating that the plasmon peak at $\sim 13\,{\rm eV}$ corresponds indeed to zero values of the real part of the macroscopic dielectric function where the imaginary part is small and showing that the crystal local-field effects have a considerable impact on the plasmon energy at small wave vectors which gives rise to an interplay with the XC effects built into the many-body kernel.

\begin{figure}[htb]
\includegraphics*[width=0.95\linewidth]{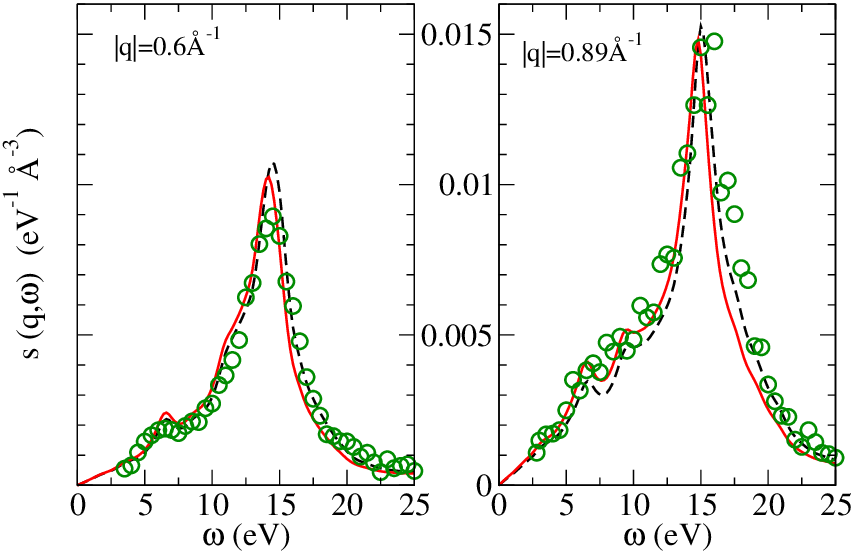}
\caption{\label{scixs06}
(Color online) Reduced dynamical structure factor $s({\bf q},\omega)=S({\bf q},\omega)/(\hbar\Omega)$ of Sc, as obtained in the RPA (dashed black line), ALDA (solid red line), and by IXS measurements (green circles). The wave vector has been taken to be {\bf q}= 8/16(001) (left panel) and {\bf q}= 12/16(001) (right panel)
in units of 2$\pi/c$, $c$ being the lattice constant in the $z$-direction.}
\end{figure}

In Fig.~\ref{scixs06}, we display new first-principles calculations (dashed black and solid red lines) and IXS measurements (green circles) of the dynamical structure factor $s({\bf q},\omega)$ of Sc at wave vectors of magnitude
$|{\bf q}|$\,=\,0.6~$\mathrm\AA^{-1}$ (left panel) and $|{\bf q}|$\,=\,0.89~$\mathrm\AA^{-1}$ (right panel)
and for energies below the $\sim 28\,{\rm eV}$ semicore onset energy (M-edge) of this material. At these wave vectors, the dynamical structure factor displays a broad plasmon peak at
$\sim 15\,{\rm eV}$, which corresponds roughly to the expected plasmon energy of
three valence electrons per atom in Sc ($r_s=2.38$);\footnote{The electron density $n_0$ of a homogeneous electron gas is typically characterized by the density parameter $r_s=(3/4\pi n_0)^{1/3}/a_0$, $a_0$ being the Bohr radius.} nevertheless, the shape of this plasmon peak, which mixes with the fine structure caused by $d$-electron transitions, bears little resemblance to the plasmon peak of a homogeneous electron gas. In particular, some structure is present at energies between 5 and 10 eV; this structure, which is also present in the imaginary part of the dielectric matrix and has also been predicted to exist for smaller wave vectors,\cite{schone02,gurtuhcp04} occurs in a region where interband transitions from the valence bands to unoccupied $p$ states above the Fermi level take place.

Fig.~\ref{scixs06} also shows that the agreement between our {\it ab initio} calculations and the experimental measurements is excellent,
both in the energy of the plasmon and in the overall shape of the energy-loss spectrum.
Furthermore, for energies below the plasmon peak the agreement between
ALDA calculations (solid red line) and IXS measurements is remarkable.

\begin{figure}[htb]
\includegraphics*[width=0.95\linewidth]{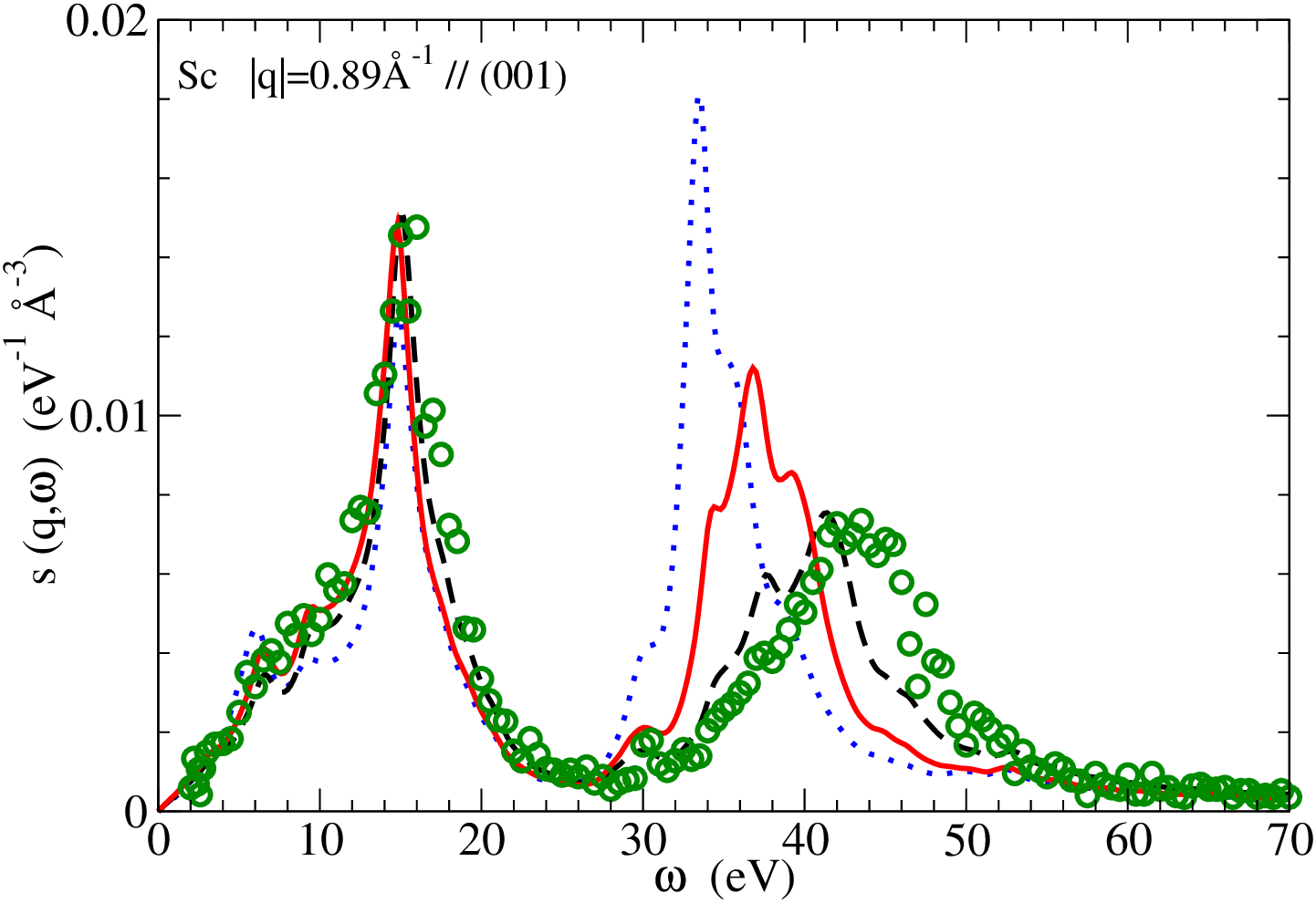}
\caption{\label{scixs089long}
(Color online) As in Fig.~\ref{scixs06}, but for {\bf q}= 12/16(001) (in units of 2$\pi/c$) and energies that include the excitation of 3$p$ semicore states. The blue dotted line denotes the diagonal (i.e. without CLFE) RPA calculation.}
\end{figure}

First-principles calculations (dashed black and solid red lines) and IXS measurements (green circles) of the dynamical structure
factor of Sc at $|{\bf q}|$\,=\,0.89~$\mathrm\AA^{-1}$ are exhibited in Fig.~\ref{scixs089long} for
energies up to $70\,{\rm eV}$, well above the $\sim 28\,{\rm eV}$ semicore onset
energy (M-edge). For comparison, diagonal RPA calculations are also plotted in the
same figure (blue dotted line), showing that the impact of crystal local-field effects is
very large just above the M-edge, where transitions from the occupied
Sc semicore $3p$ states to the lowest conduction bands occur. Crystal
local-field effects bring the onset of the M-edge to larger energies and
reduce the corresponding peak height, in close agreement with experiment.
The shift to larger energies stems from the fact that the semicore
states involved in these transitions are highly localized in real space 
and therefore  neglecting  the off-diagonal terms in Eq.~(\ref{eq9}),
 and subsequently in Eq.~(\ref{eps}), 
underestimates the large dynamical screening effects near the semicore edge.

Figure ~\ref{scixs089long} shows, however, that although our full RPA calculation (black dashed  line) is in very good agreement with experiment (green circles) there is a mismatch between the ALDA calculation (red solid line) and the measured spectrum. This discrepancy calls for
further investigations of (i) an improved description of single-particle
energies beyond the LDA and (ii) accurate many-body kernels accounting for XC effects beyond the ALDA. Both RPA and ALDA calculations have been carried out with the very same form of the XC potential entering the KS equation of DFT. Hence, we can rule out our approximate XC potential as the origin of the
{\it large} discrepancy between ALDA calculations and experiment, and sources of error of our ALDA calculations should be searched both in the adiabaticity and the locality of the ALDA XC kernel in a solid where inhomogeneities of both the static unperturbed electron density and the induced density play an important role, especially when semicore states with large binding energies are involved.

\begin{figure}
\includegraphics*[width=0.95\linewidth]{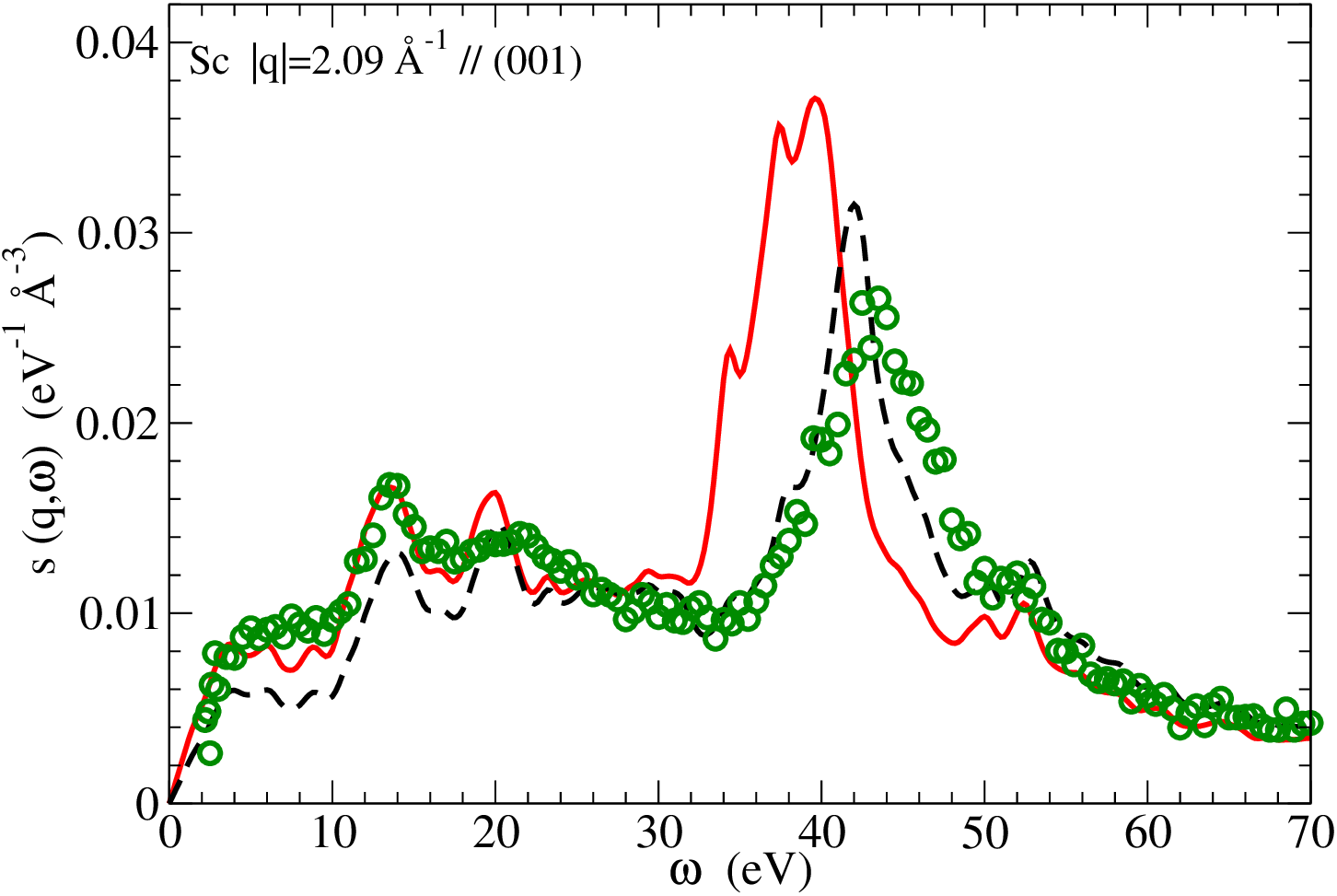} \\
\vspace*{1cm}
\includegraphics*[width=0.95\linewidth]{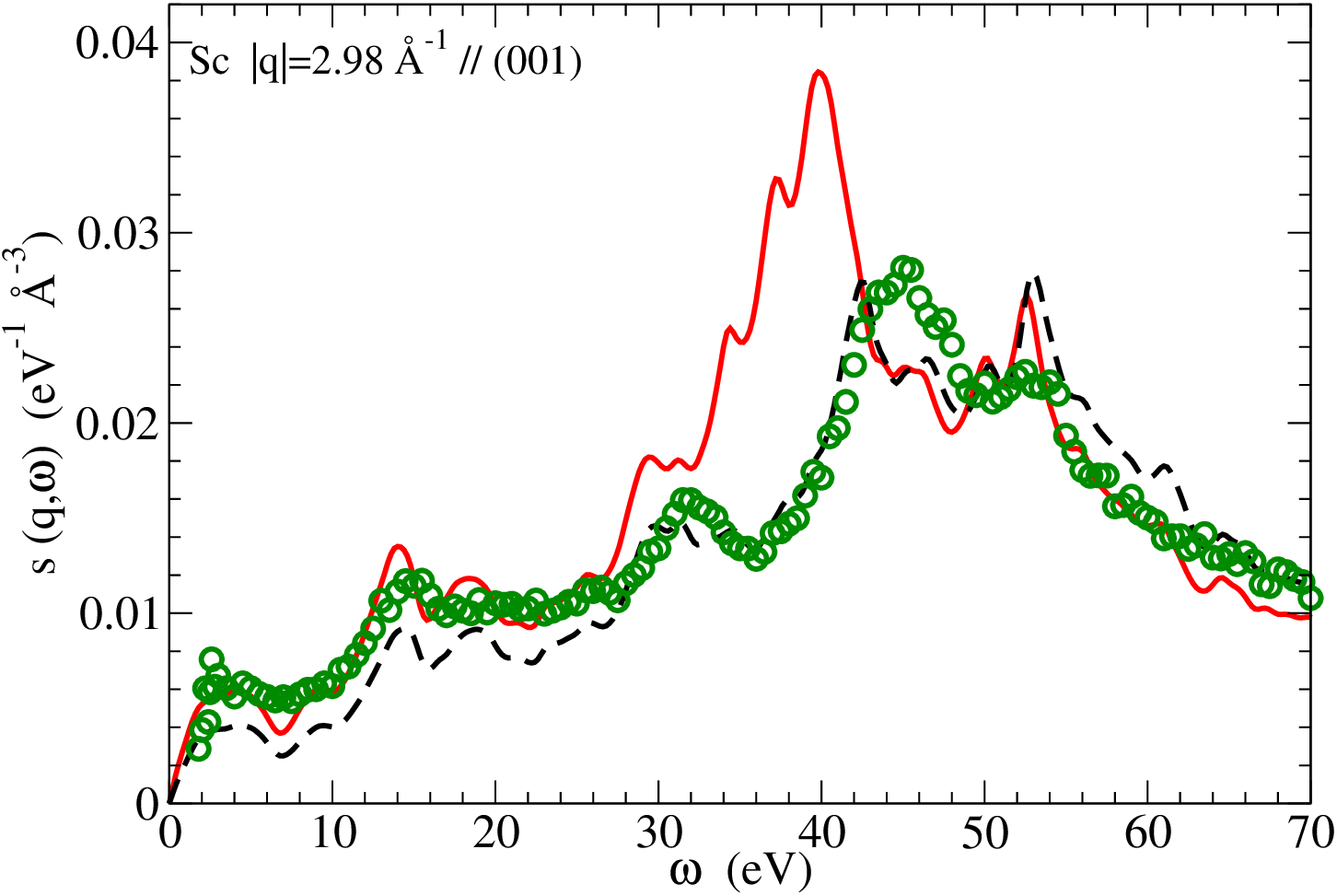}
\caption{\label{scixsq2}
(Color online)
Dynamical structure factor $s({\bf q},\omega)$ of Sc, as obtained in the RPA (black dashed line), ALDA (red solid line), and by IXS measurements (green circles). The wave vector has been taken to be {\bf q}= [4/16(001)+(002)] (upper panel) and {\bf q}= [8/16(001)+(002)] (lower panel) in units of 2$\pi/c$, $c$ being the lattice constant in the $z$-direction.}
\end{figure}

In Fig.~\ref{scixsq2}, we show calculations and measurements of the dynamical structure
factor of Sc, as in Fig. 2, but now at wave vectors of magnitude $|{\bf
q}|$\,=\,2.09~$\mathrm\AA^{-1}$ (upper panel) and $|{\bf
q}|$\,=\,2.98~$\mathrm\AA^{-1}$ (lower panel). At these large wave vectors, the
plasmon energy falls into the electron-hole pair excitation continuum so there is no well-defined plasmon peak. Instead, a new feature appears in
the dynamical structure factor at $\sim 4\,{\rm eV}$, which is
very sensitive to crystal local-field effects. A close examination of the various
transitions that contribute to the KS density-response function of Eq.~(\ref{eq9}) indicates
that this new feature, which is absent for small wave vectors, is originated in the
presence of $d$-to-$d$ transitions involving narrow $d$ complexes lying
below and above the Fermi level. Such direct $d$-to-$d$ excitations 
have not been observed in
optical absorption measurements (where the wave vector is zero)
and EELS experiments (where the wave vector is small),
 which
underscores the importance of IXS for the examination of the electronic
structure of these materials.

An inspection of Fig.~\ref{scixsq2} also shows that introducing many-body XC effects within
the ALDA (red solid line) leads to a faithful description of the low-energy part of the
measured energy-loss spectrum (green circles), as occurs for smaller wave vectors (see Figs.~\ref{scixs06}
and \ref{scixs089long}), even at $\sim 4\,{\rm eV}$ where the energy loss is dictated by the presence of $d$-to-$d$ transitions. At energies above the M-edge, the agreement between our full RPA calculations (black dashed line) and the IXS measurements is excellent, but there is a considerable mismatch between the ALDA calculation (red solid line) and the measured spectrum, as occurs for smaller wave vectors (see Fig.~\ref{scixs089long}).

A closer look into the semicore energy region in Figs.~\ref{scixs089long} and 
~\ref{scixsq2} still shows a 2-3~eV mismatch between the calculated RPA and the measured IXS semicore excitation onset. This stems from a well-known
deficiency of the local-density approximation we have used to describe 
our single-particle orbitals,\footnote{This deficiency contributes to the discrepancy between ALDA calculations and experiment; however, the {\it large} discrepancies  that are visible in Figs.~\ref{scixs089long} and 
~\ref{scixsq2} should be searched in the adiabaticity and the locality of the ALDA XC kernel.}  which in general underbinds the energy of highly 
localized occupied states and is known to occur even in the case of a simple metal such as Al.\footnote{We have carried out RPA calculations and IXS measurements of the dynamical structure factor of Al, and we have found that the LDA predicts the 2$p$ semicore states of this material to be located at
$\sim$65~eV while the IXS measurements show the 2$p$ semicore excitation onset at  $\sim$72~eV.}

\subsection{Cr}

Cr ([Ar] 3$d^5$ 4$s^1$) lying in the middle of the $3d$ period exhibits a high density of both occupied and unoccupied $d$ states near the Fermi level. Hence, this transition metal is the ideal candidate for the investigation of the characteristic $d$-to-$d$ transitions that we have observed in the case of Sc.  

\begin{figure}[htb]
\includegraphics*[width=0.95\linewidth]{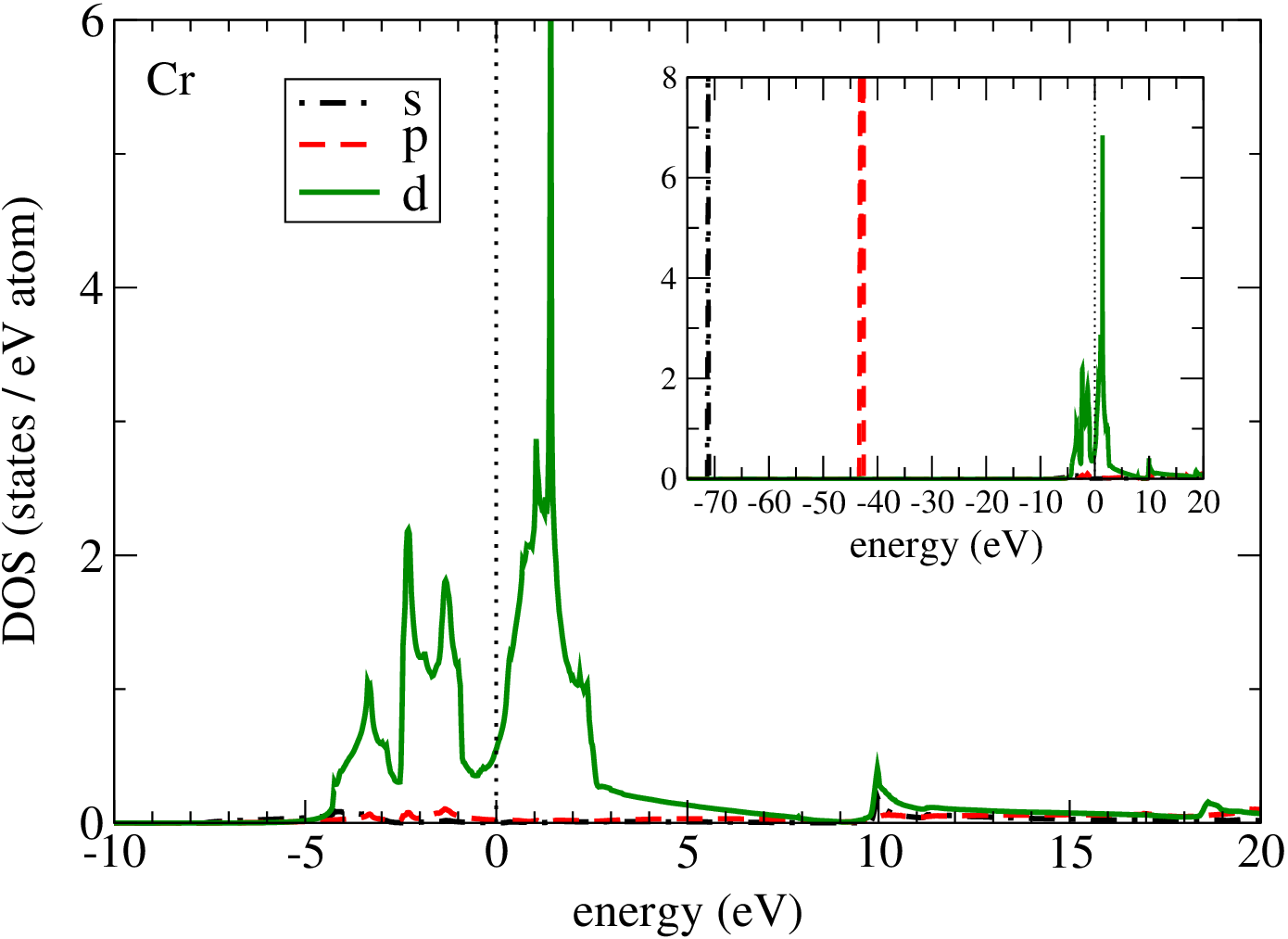}
\caption{\label{doscr}
(Color online) Density of states (DOS) of Cr near the Fermi level (dotted line)
for each component of the angular momentum inside the atomic sphere.
Black dash-dotted, red dashed,  and green solid
 lines represent the number of states per atom and per eV
 of $s$, $p$, and $d$ character, respectively.
The inset shows a wider energy range,
in which the 3$s$ (black dash-dotted line) and  3$p$ (red dashed line) semicore states are visible.}
\end{figure}

The density of states of Cr is depicted in Fig.~\ref{doscr}, where 
$d$ states distributed at both sides of the Fermi level are represented
by green solid lines. As in Fig.~\ref{dosscti}, the inset of this figure shows the location of the 3$s$ (black dash-dotted line) and 3$p$ (red dashed line) semicore states at $\sim$70~eV and $\sim$42~eV below the Fermi level, which hold 2 and 6 electrons per atom, respectively.   

Inelastic x-ray scattering measurements of the energy-loss spectra of Cr were reported recently by Montano and Macrander\cite{montano00} for momentum transfers ranging from 0.80 to 5.0$\mathrm\AA^{-1}$ along the (110) direction. Spectral features were found corresponding to plasmon excitations, as well as a second peak corresponding to excitations from the $3p$ semicore states to the $3d$ valence states; however, no structure was visible at low energies near $\sim 4$ eV, where the x-ray spectrum was dominated by the quasielastic phonon scattering feature.  To the best of our knowledge, no theoretical investigations
of the dynamical response  of this material have been reported before.

\begin{figure}[htb]
\includegraphics*[width=0.95\linewidth]{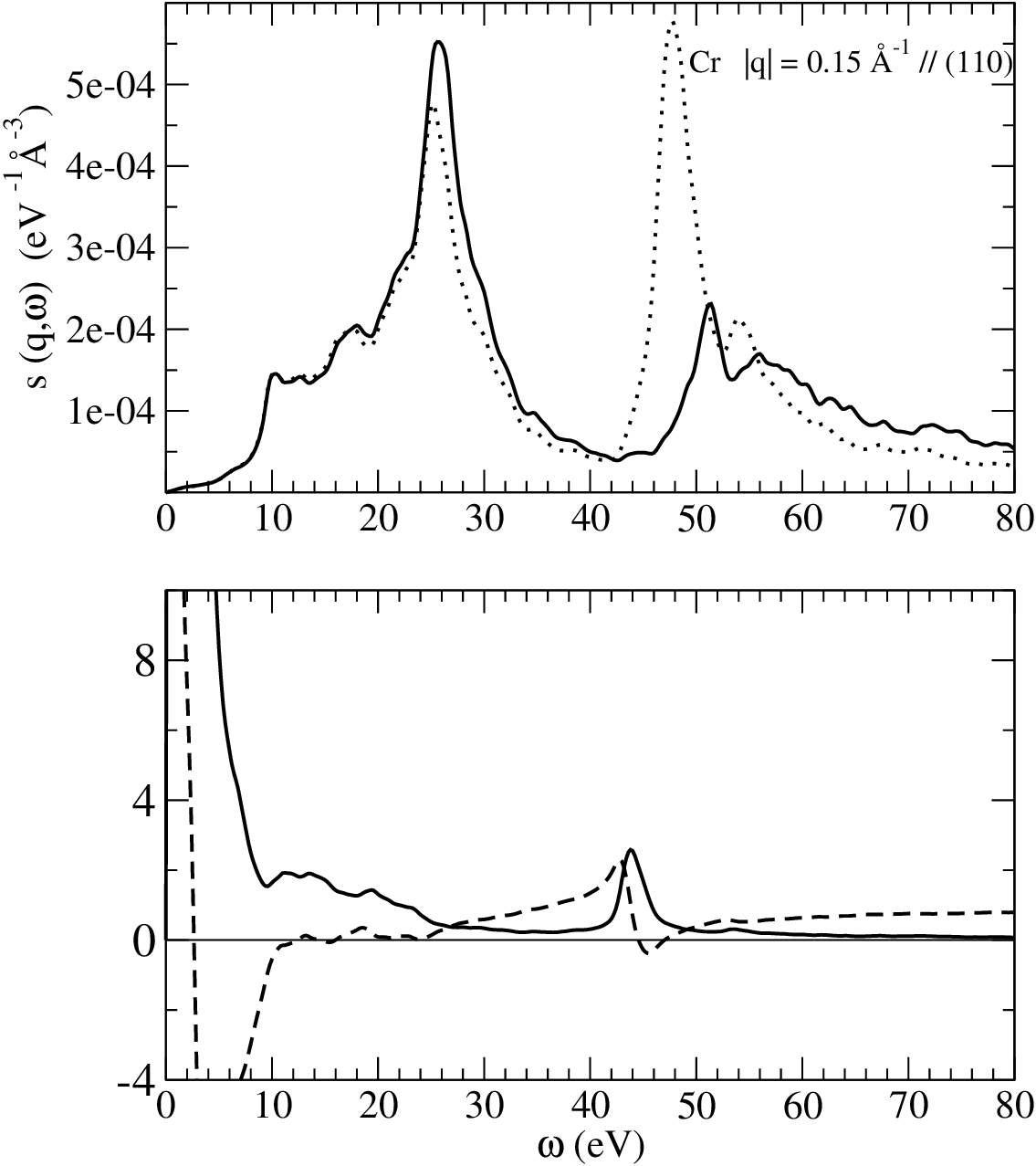}
\caption{\label{cr015}
Top panel: RPA full (solid line) and diagonal (dotted line) calculations of the dynamical structure factor $s({\bf q},\omega)$ of Cr.
Bottom panel: RPA calculations of the real (dashed line) and imaginary (solid line) parts of the first diagonal element of the dielectric matrix $\epsilon_{{\bf G},{\bf G}'}({\bf q},\omega)$ of Cr. In both panels, the wave vector has been taken to be {\bf q}= 1/10(110) in units of 2$\pi/a$, $a$ being the lattice parameter.}
\end{figure}

The top panel of Fig.~\ref{cr015} shows first-principles RPA diagonal (dotted line) and full (solid line) calculations of the dynamical structure factor
$s({\bf q},\omega)$ of Cr for small wave vectors of magnitude $|{\bf q}|\,=\,0.15\,\mathrm{\AA^{-1}}$ along the (110) direction. At energies below the $\sim 43\,{\rm eV}$ semicore onset energy (M-edge), where crystal local-field effects are found to be small (diagonal and full calculations are very close from each other), we find a prominent broad feature at
$\sim 26\,{\rm eV}$, which agrees with the expected plasmon energy of 6 valence electrons per atom in Cr ($r_s=1.48$) and is clearly identifiable as a collective excitation of hybridized $3d^5$ and $4s^1$ valence electrons. Indeed, this peak corresponds (see the bottom panel of Fig.~\ref{cr015}) to a zero value of the real part of the matrix element $\epsilon_{0,0}({\bf q},\omega)$ where ${\rm Im}\epsilon_{0,0}({\bf q},\omega)$ is small.\footnote{At the wave vectors and energies under consideration crystal local-field effects in Cr are small and one finds, therefore, $\epsilon_M({\bf q},\omega)\approx\epsilon_{0,0}({\bf q},\omega)$.}

At energies above the M-edge, transitions from the localized Cr semicore $3p$ states below the Fermi level to the lowest conduction bands take place. As a result, the impact of crystal local-field effects is large (see the top panel of Fig.~\ref{cr015}), as occurs in the case of Sc. These effects bring the onset of the M-edge to larger energies and considerably reduce the peak height above the M-edge, which becomes a broad feature around 50-60 eV involving primarily excitations out of the $3p$ band of Cr.

\begin{figure}[htb]
\includegraphics*[width=0.95\linewidth]{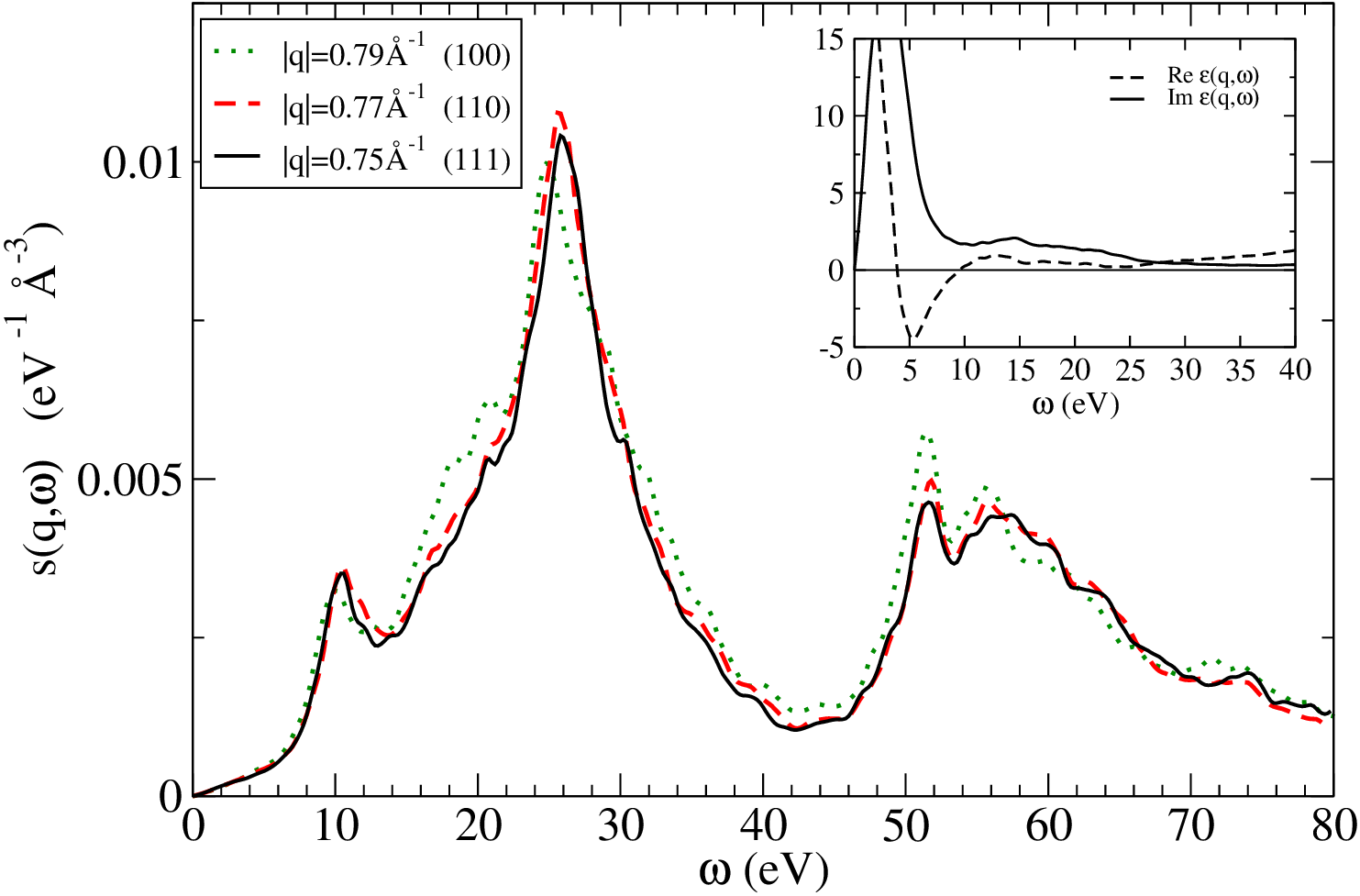}
\caption{\label{cr07isot}
(Color online)
RPA calculations of the dynamical structure factor $s({\bf q},\omega)$ of Cr at wave vectors of magnitude $|{\bf q}|\approx 0.8\,\mathrm{\AA^{-1}}$ along three high-symmetry directions.
Dotted green, dashed red, and solid black lines correspond to 
{\bf q}= 4/11(100), {\bf q}= 5/20(110), and {\bf q}= 2/10(111)
wave vectors, respectively, in units of 2$\pi/a$, $a$ being the lattice parameter.
The inset shows the real (dashed line) and imaginary (solid line) parts of the first diagonal element of the RPA dielectric matrix $\epsilon_{{\bf G},{\bf G}'}({\bf q},\omega)$ at $|{\bf q}|\,=\,0.75{\mathrm\AA^{-1}}$.}
\end{figure}

In order to investigate the dependence of the dynamical structure factor of Cr on the direction of the wave vector, we have plotted in Fig.~\ref{cr07isot} the RPA dynamical structure factor $s({\bf q},\omega)$ of Cr at wave vectors of magnitude $|{\bf q}|\approx 0.8\,\mathrm{\AA^{-1}}$ along the (100), (110), and (111) directions.\footnote{In order
to be able to choose similar values of the magnitude of the 
momentum transfer along the three directions under study, we have used
two different samplings of the BZ, 
$10\times10\times10$ and $11\times11\times11$. As the damping parameter $\eta$ has been taken to be $\eta\,=\,0.65$~eV in both cases, small differences among the three curves may be due not only to the fact that the magnitude of the wave vector is not exactly the same but also to the fact that different samplings of the BZ have been used with the same value of $\eta$.} This figure shows that the dynamical structure factor of Cr is rather similar for all high-symmetry propagation directions. In particular, we note that the long-wavelength collective excitation at $\sim 26$~eV (see Fig.~\ref{cr015}) is still visible along all propagation directions when the magnitude of the wave vector is $|{\bf q}|\approx 0.8\,\mathrm{\AA^{-1}}$. This is in agreement with the IXS measurements reported by Montano and Macrander\cite{montano00} for this magnitude of the wave vector along the (110) direction. Furthermore, that this feature is still originated in the building up of a collective excitation follows from the inset of Fig.~\ref{cr07isot}, where the real and imaginary parts of the matrix element $\epsilon_{0,0}({\bf q},\omega)$ are displayed: this feature corresponds to a zero value of ${\rm Re}\epsilon_{0,0}({\bf q},\omega)$ (dashed line) where ${\rm Im}\epsilon_{0,0}({\bf q},\omega)$ (solid line) is small. On the other hand, a comparison of the high-energy part of the energy-loss spectra of Figs.~\ref{cr015} and \ref{cr07isot} indicates that the high-energy broad feature around 50-60 eV, which is also rather insensitive to the propagation direction and which involves primarily excitations out of the $3p$ band of Cr, has a considerable wave-vector dependence, becoming more pronounced as the wave vector increases.

\begin{figure}[htb]
\includegraphics*[width=0.95\linewidth]{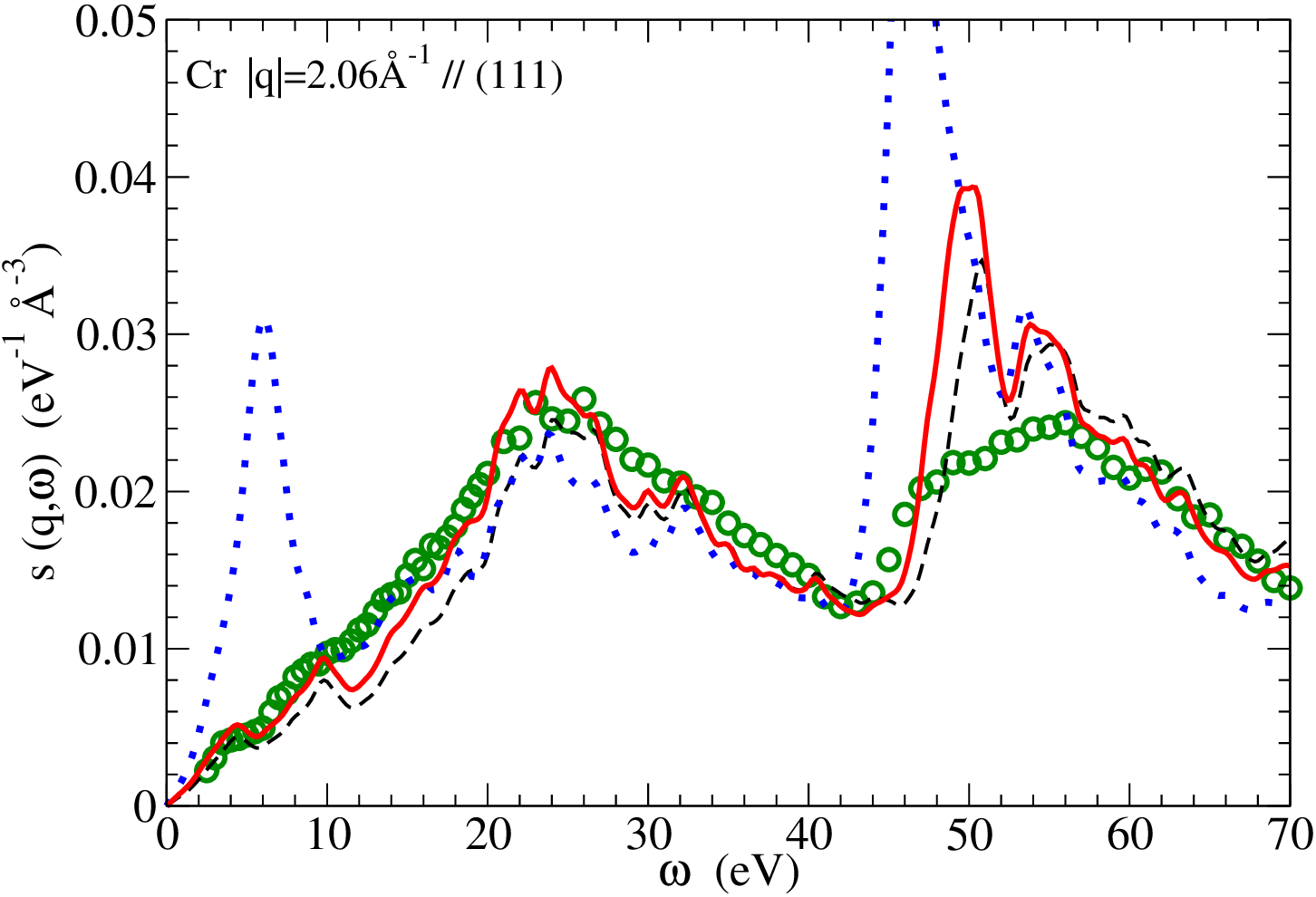} \\
\vspace*{1cm}
\includegraphics*[width=0.95\linewidth]{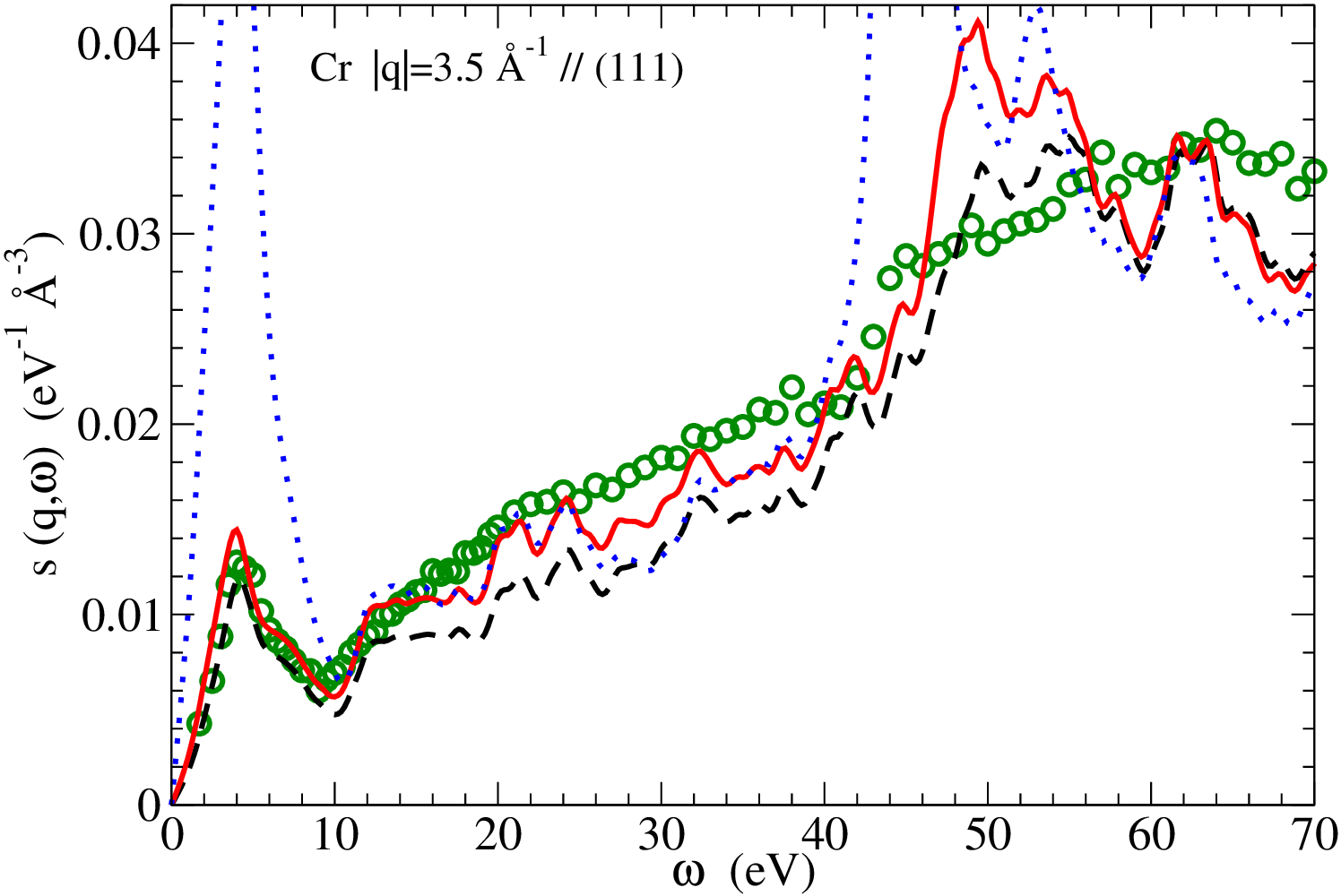}
\caption{\label{crALDA}
(Color online)
Dynamical structure factor of Cr (CLFE, i.e. off-diagonal terms in Eq.(\ref{eps}),
 included) along the  (111) direction for
wave vectors
{\bf q }= 6/11(001) (top panel) and {\bf q }= 10/11(001) (bottom panel),
within RPA (black dashed line) and ALDA (red solid line).
The blue dotted line represents the calculation without CLFE and green circles,
IXS measurements.}
\end{figure}

First-principles calculations and IXS measurements of the dynamical structure factor $s({\bf q},\omega)$ of Cr are displayed in Fig.~\ref{crALDA} at large wave vectors of magnitude $|{\bf q}|\,=\,2.06{\mathrm\AA^{-1}}$ and 
$|{\bf q}|\,=\,3.5{\mathrm\AA^{-1}}$ along the (111) direction. As in the case of Sc, at these large wave vectors the prominent broad peak of Figs.~{\ref{cr015} and \ref{cr07isot} at $\sim 26$ eV, which is in the nature of collective excitations, is now replaced by a less pronounced broad feature at energies below the M-edge originated from single electron-hole excitations, and the high-energy feature above the M-edge is found to gain weight as the wave vector increases.

For small wave vectors, crystal local-field effects only play a role at energies above the M-edge (see Figs.~{\ref{cr015} and \ref{cr07isot}), where the energy-loss spectrum is dominated by transitions from very localized $3p$ bands below the Fermi level. 
The striking feature of Fig.~\ref{crALDA} is precisely the fact that as the wave vector increases huge crystal local-field effects are present not only at energies above the M-edge but also at very low energies  below 10 eV. In the absence of crystal local-field effects (blue dotted line), the energy loss spectra at
 the large wave vectors under consideration are clearly dominated by
 two distinct features at 4-6 eV and 40-60 eV. A close examination of the
 various transitions that contribute to the KS density-response function 
of Eq.~(\ref{eq9}) indicates that these features involve excitations to
 unoccupied $d$ states above the Fermi level coming primarily 
from occupied $d$ and $p$ states, respectively. 
As in the case of Sc, high-energy transitions from occupied $p$ states 
occur for arbitrary values of the wave vector; however, direct $d$-to-$d$ transitions are not observed for small wave vectors.
In the presence of crystal local-field effects (dashed black and solid red lines), both peak heights are considerably reduced and our calculated low-energy feature is brought into excellent agreement with experiment. Both theory and experiment exhibit the presence of a very well defined low-energy peak at $\sim 4$ eV, which is originated in the presence of direct $d$-to-$d$ transitions and had never been observed before.

An inspection of Fig.~\ref{crALDA} clearly indicates that (i) first-principles RPA calculations provide a reasonably good description of the measured low-energy peak at $\sim 4$ eV, provided that crystal local-field effects are included, (ii) inclusion of XC effects within the ALDA brings the low-energy peak at $\sim 4$ eV into excellent agreement with experiment, and (iii) the ALDA, which accurately accounts for XC effects at low energies even at large wave vectors, fails to give a quantitative description of the measured energy-loss spectra at energies above the M-edge. This suggests that for an accurate understanding of the energy-loss spectra of transition metals in the whole energy-range under study one should go beyond and adiabatic local-density approximation in the description of XC effects not present in the RPA. We also note that as in the case of Sc (see Figs.~\ref{scixs089long} and
\ref{scixsq2}) both RPA (black dashed  line) and ALDA (red solid line) calculations predict an energy onset for the excitation of $3p$ electrons in Cr that is too large, which should be a consequence of the failure of the LDA to describe the bonding energies of the semicore KS states for this material. 

\begin{figure}[htb]
\includegraphics*[width=0.95\linewidth]{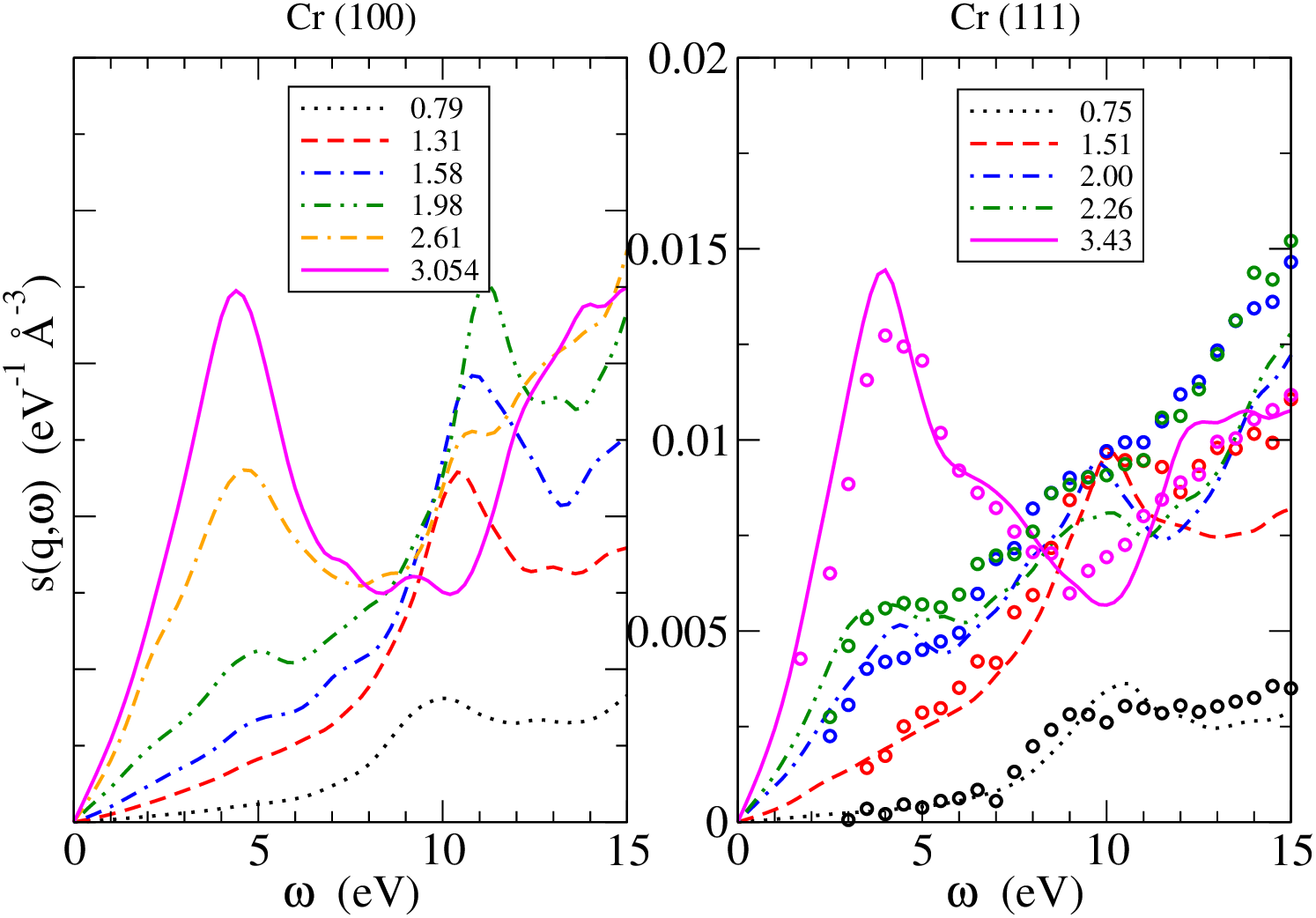}
\caption{\label
{cr4eV}
(Color online)
Formation of the $d\,\rightarrow\,d$ transition peak along the 
 (100) and (111) directions,
for different values of the wave vector (shown in ${\mathrm \AA^{-1}}$).}
\end{figure}

The $d$-to-$d$ low-energy peak that is visible at $\sim 4$ eV in Fig.~\ref{crALDA} is absent for small wave vectors and is therefore expected to have a very strong wave-vector dependence. Hence, in Fig.~\ref{cr4eV} we have illustrated this wave-vector dependence by plotting our full ALDA calculations (lines) and IXS measurements (circles) of the dynamical structure factor
$s({\bf q},\omega)$ of Cr at wave vectors of magnitude ranging from $\sim 0.7$ to 3.5${\mathrm \AA^{-1}}$ along the (100) and (111) directions (left and right panels, respectively). First, we note that both theory and experiment indicate that the weight of the $d$-to-$d$ low-energy peak at $\sim 4$ eV, which is absent for small wave vectors, considerably increases with ${\bf q}$. Secondly, it is clear from this figure that ALDA calculations of this low-energy feature are in remarkable agreement with experiment for {\it all} wave vectors and prove not only the presence of considerable crystal local-field effects below the M-edge but also the importance of IXS for the examination of the electronic structure of a $3d$ transition metal such as Cr.  

Finally, we note that the $d$-to-$d$ low-energy peak that is visible at $\sim 4$ eV in Figs.~\ref{scixsq2} and \ref{crALDA} is considerably sharper and more intense in the case of Cr (Fig.~\ref{crALDA}) than in the case of Sc (Fig.~\ref{scixsq2}). This is simply due to the fact that in the case of Cr the density of $d$ states is considerably large not only above the Fermi level but also below, thus increasing the number of available $d$-to-$d$ transitions. 

\subsection{The $3d$ series}

\begin{figure}[htb]
\includegraphics*[width=0.93\linewidth]{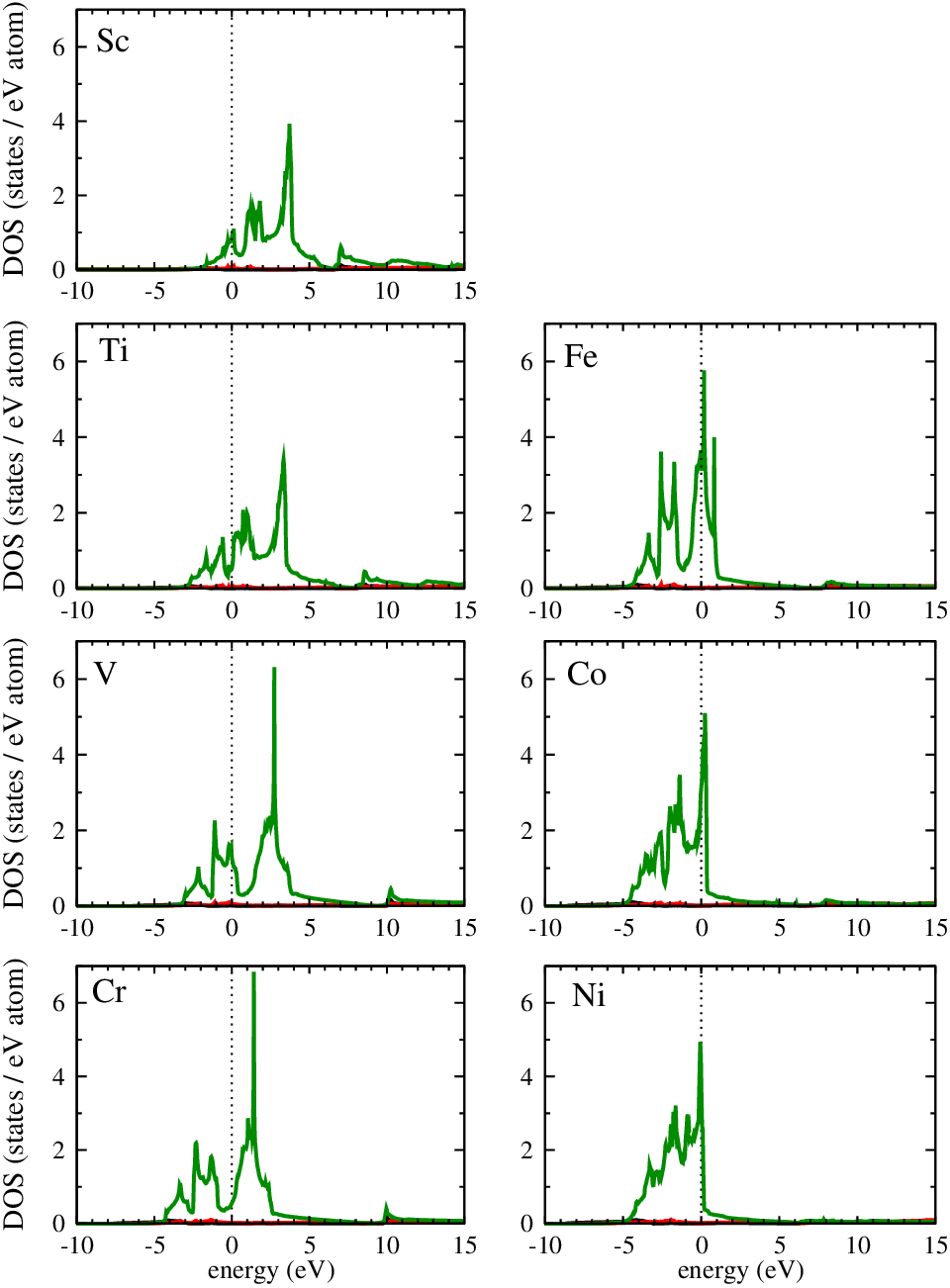} \\
\caption{\label{dosall3d} (Color online) $d$ content of the density of states in the  3$d$
transition metal row, as we move from Sc to Ni (in the figure from
top to bottom and left to right).}
\end{figure}

As we move along the 3$d$ transition-metal period, from Sc to Ni,
the 3$d$ shell fills up causing a shift of the 3$d$ bands towards
the occupied side of the band structure below the Fermi level, as illustrated by the density of states that have been plotted in Fig.~\ref{dosall3d}. Here we focus on the impact that this filling up of the $3d$ bands has on the collective excitations that are present for small wave vectors and, more importantly, on the {\it new} low-energy feature that is originated in the building up of $d$-to-$d$ transitions.

\begin{widetext}
\begin{center}
\begin{table}[h!]
\begin{tabular}{l| c c c c c c c }
                	& Sc & Ti & V & Cr & Fe & Co & Ni  \\
 \hline
Electronic structure [Ar]& \quad$3d^14s^2$\qquad & \quad$3d^24s^2$\quad & \quad$3d^34s^2$\quad & \quad$3d^54s^1$\quad & \quad$3d^64s^2$\quad & \quad$3d^74s^2$\quad & \quad$3d^84s^2$\quad \\
Crystal structure & hcp & hcp & bcc & bcc  & bcc  & hcp &fcc \\
Points in  IBZ & 90& 90& 47 & 47 & 47 & 90 & 72 \\
low {\bf q}  (${\mathrm \AA^{-1}}$) & 0.44& 0.42 & 0.41 & 0.43 & 0.44 & 0.37 & 0.29 \\
large {\bf q}  (${\mathrm \AA^{-1}}$) & 2.98& 3.01 & 2.91 & 3.05 & 3.06 & 2.99 & 3.27 \\
\hline\hline
\end{tabular}
\caption{\label{3dtable}
Parameters used for the comparative study of the 3$d$ transition metals. Low {\bf q} and large  {\bf q} refer to the magnitude of the wave vector
used in Figs.~\ref{sqwall3dlowq} and \ref{sqwall3d4eV}, respectively.}
\end{table}
\end{center}
\end{widetext}

\subsubsection{Collective excitations}

\begin{figure}[htb]
\includegraphics*[width=0.85\linewidth]{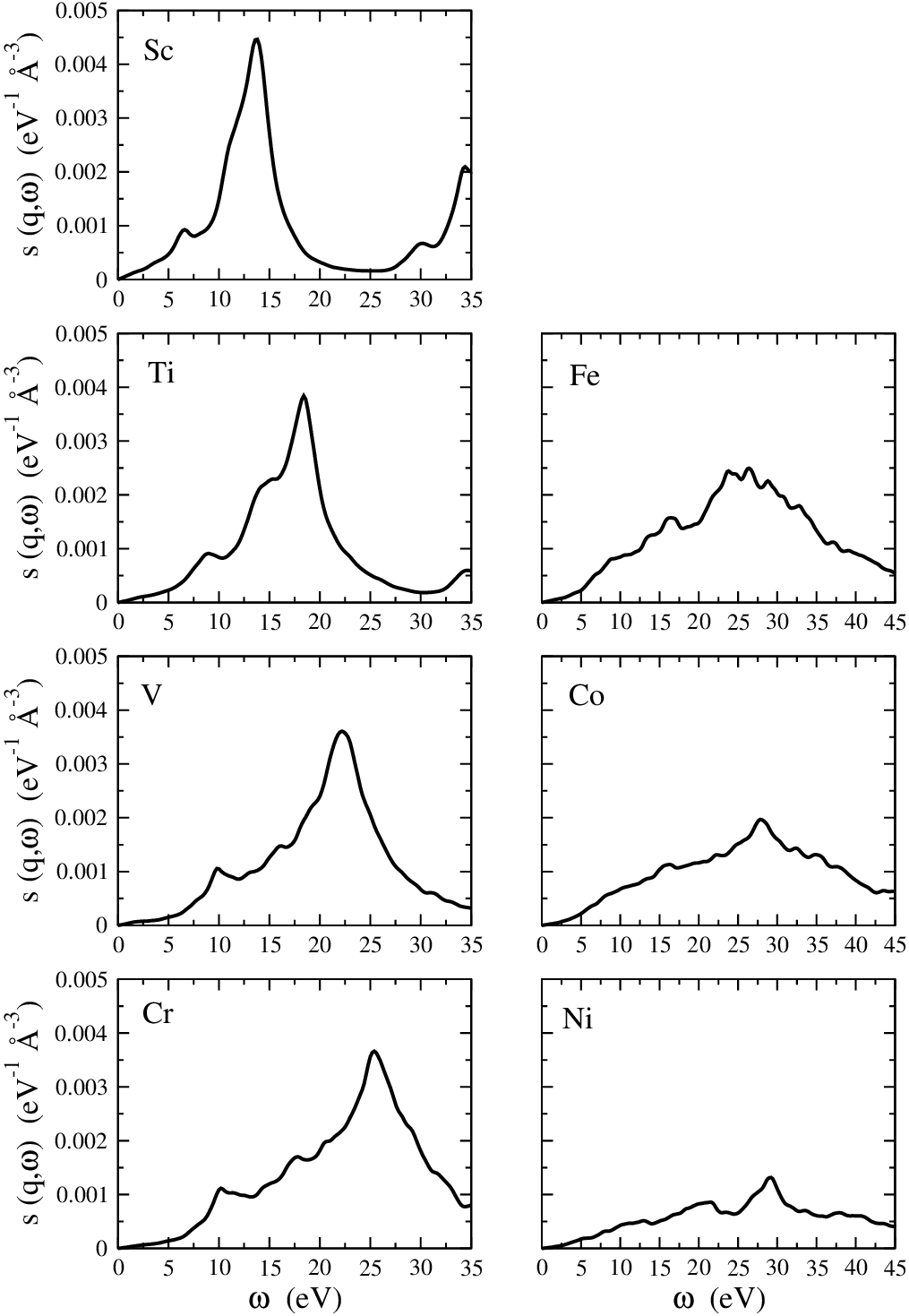} \\
\caption{\label{sqwall3dlowq}
Dynamical structure factor of the 3$d$ transition metals along the (001) direction
for a  similar magnitude of the momentum transfer 
($|{\bf q}|\,\sim\,0.4$~${\mathrm \AA^{-1}}$)
and  energies below the M-edge.
The exact value  of the momentum transfer can be found in Table~\ref{3dtable}.}
\end{figure}

Full ALDA calculations of the dynamical structure factor $s({\bf q,\omega})$ of $3d$ transition metals are depicted in 
Fig.~\ref{sqwall3dlowq} at a low wave vector of magnitude 
$\sim\,0.4$~${\mathrm \AA^{-1}}$ along the (001) direction and for 
energies below the M-edge.
This figure clearly shows that in the case of $3d$ metals with five 
or fewer $d$ states occupied per atom (Sc, Ti,
V, and Cr) a broad peak can be identified, which mixes with the fine structure due to $d$-electron transitions. As the $3d$ band fills up (Fe, Co, and Ni), however, the number of states available for real transitions from occupied $d$ bands below to unoccupied $p$ states above the Fermi level considerably increases. Furthermore, the combination of band-structure effects and the building-up of collective modes of $d$ electrons yields broader structures in these materials which may still be in the nature of collective excitations but which are not well defined any more.

We note in passing that our first-principles calculations are in good agreement with the IXS measurements of the dynamical structure factor of Ti 
reported by Macrander {\it et al.}.\cite{macrander96}

\subsubsection{$d\,\rightarrow\,d$ transitions}

\begin{figure}[htb]
\includegraphics*[width=0.75\linewidth]{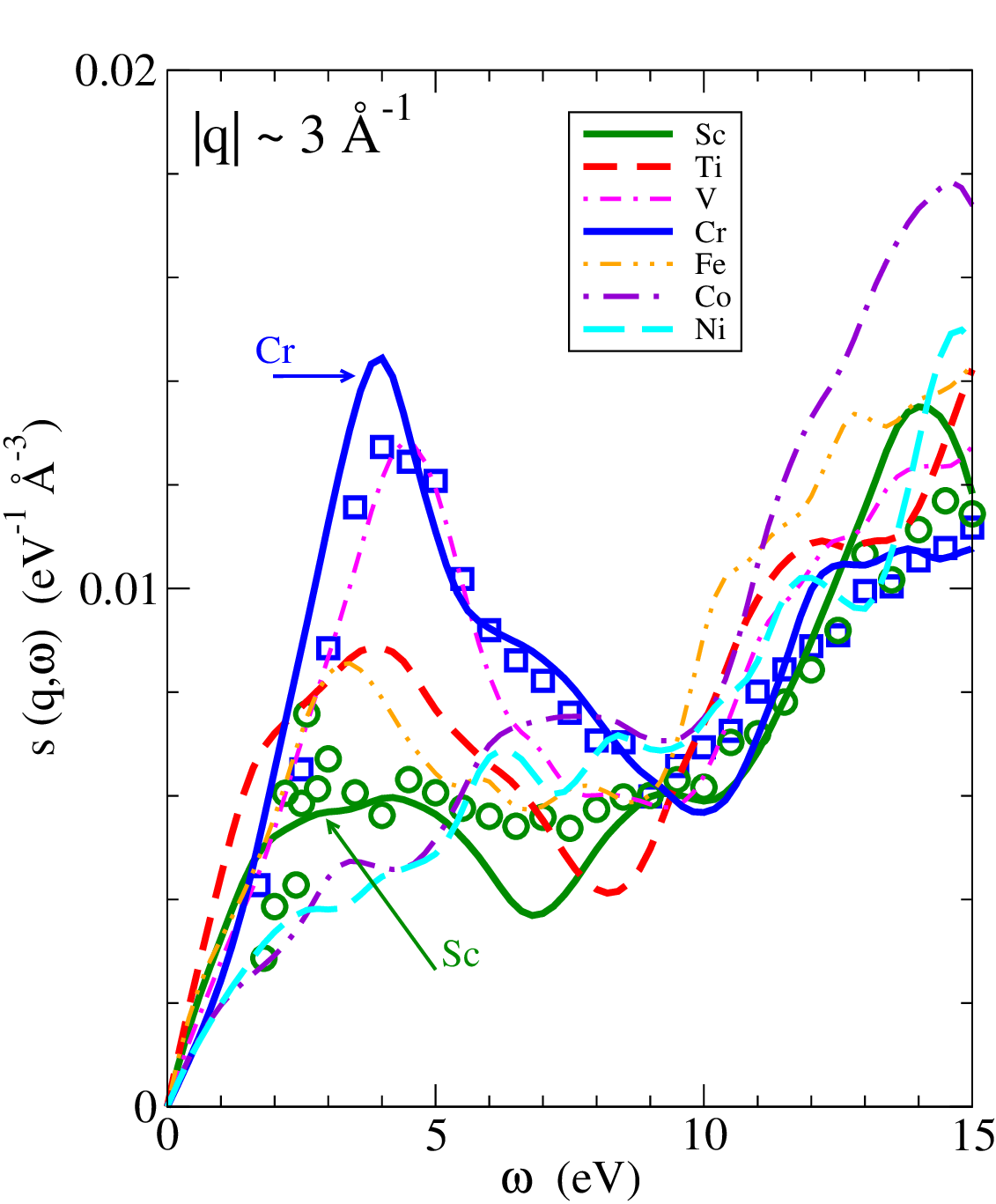}
\caption{\label
{sqwall3d4eV}
(Color online)
Low energy peak arising from $d\,\rightarrow\,d$ transitions
for a large wave vector,
as we move from left (Sc) to right (Ni) along the
3$d$ row.
The magnitude of the momentum transfer for each metal can be found in
Table~\ref{3dtable}. 
Lines denote the {\it ab initio} ALDA calculations and circles and squares  
correspond to the normalized IXS measurements on Sc and Cr, respectively.}
\end{figure}

Figure~\ref{sqwall3d4eV} shows full ALDA calculations of the dynamical structure factor $s({\bf q,\omega})$ of the $3d$ transition metals Sc, Ti, V, Cr, Fe, Co, and Ni and the corresponding IXS measurements for the $3d$ transition metals Sc (green circles) and Cr (blue squares) at a large wave vector of magnitude $|{\bf q}|\sim$3~${\mathrm \AA^{-1}}$ and low energies well below the M-edge. This figure clearly shows the presence of a low-energy peak at $\sim 4$ eV, which follows the trends of the density of states depicted in Fig.~\ref{dosall3d} as should be expected from a feature that is originated in the building up of $d$-to-$d$ transitions involving narrow $d$ complexes lying below and above the Fermi level. The weight of this characteristic peak is small in Sc, which holds one single occupied $d$ state per atom, it considerably increases in the case of Ti and V, as more $d$ states become occupied, and it is extremely well defined in the case of Cr, with a half-filled $3d$ band below the Fermi level. As still more $d$ states become occupied, as occurs in the case of Fe and Co, the number of states available for real transitions involving $d$-to-$d$ transitions decreases again, thereby decreasing the weight of the $\sim 4$ eV peak. Finally, in the case of Ni, with only two unoccupied $3d$ states per atom, this feature is hardly visible.   

\section{Summary and conclusions}

We have presented a combined theoretical and experimental investigation of the dynamical structure factor of various $3d$ transition metals with $d$ states below and above the Fermi level. Calculations have been carried out for Sc, Ti, V, Cr, Fe, Co, and Ni by first expanding single-particle Bloch states in an LAPW basis and then computing the density-response function of the solid either in the RPA or the ALDA with full inclusion of crystal local-field effects. The experimental dynamical structure factor has been obtained for Sc and Cr by first performing nonresonant IXS measurements of the relative intensity of the scattered beam at a given wave vector and energy and then reducing the measured spectra to absolute units from sum-rule considerations.

For small wave vectors, a collective excitation of hybridized $3d$ and $4s$ electrons is observed, which is very well described by our ALDA calculations and which in the case of $3d$ metals holding five or fewer $d$ states per atom (Sc, Ti, V, and Cr) has a rather well defined energy. At large wave vectors, IXS measurements exhibit a {\it new} feature at $\sim 4$ eV, which (i) had never been observed before, (ii) we have found to be originated in the presence of $d$-to-$d$ transitions involving $d$ states below and above the Fermi level, and (iii) is remarkably well described by our ALDA calculations, as long as crystal local-field effects are fully included. This new feature is absent at the small wave vectors accessible by EELS, and underscores the importance of IXS for investigations of the electronic structure and the dynamical response of transition metals.

Our ALDA calculations and IXS measurements of the dynamical structure factor of the transition metals Sc and Cr have been found to show a remarkable agreement at energies below the M-edge. However, at energies above the M-edge significant discrepancies are found to exist between theory and experiment. Discrepancies between theory and experiment at the onset of the semicore excitations in Cr provide a signature of the single-particle LDA energies underestimating the binding energy of the semicore states. Discrepancies at the intensity and overall shape of the edge, especially in the case of Sc, should be due to the fact that the adiabatic local-density approximation for the many-body kernel breaks down at these large energies. These discrepancies call for further investigations of the accuracy of various approximations for the XC kernel of TDDFT.

\acknowledgments
 
I.G.G. and J.M.P. acknowledge partial support by the UPV/EHU, the Basque 
Unibertsitate, Hezkuntza  eta Ikerketa Saila, the MCyT, and the EC 6$^{\rm th}$ framework Network of Excellence NANOQUANTA (NMP4-CT-2004-500198). 
W.K.  acknowledges support from the U.S. DOE under Contract No.
DE-AC02-98CH10886. 
A.G.E. acknowledges support from NSF ITR DMR-0219332.  
K.D.F. acknowledges support by the National
Science Foundation and the National Institutes of Health under
award DMR-0225180.
ORNL research sponsored by the DOE, Office of Science, DMS under contract with
UT-Battelle, LLC. 
The UNICAT facility at the Advanced Photon Source (APS) is
supported by the U.S. DOE under Award No. DEFG02-91ER45439,
through the Frederick Seitz Materials Research Laboratory at the
University of Illinois at Urbana-Champaign, the Oak Ridge
National Laboratory (U.S. DOE contract DE-AC05-00OR22725 with
UT-Battelle LLC), the National Institute of Standards and
Technology (U.S. Department of Commerce) and UOP LLC. The APS is
supported by the U.S. DOE, Basic Energy Sciences, Office of
Science under contract No. W-31-109-ENG-38.
I.G.G gratefully acknowledges the hospitality of the University of Tennessee,
Knoxville,
 and Oak Ridge National Laboratory, where most of these calculations were
carried out.

\bibliography{paper}

\begin{thebibliography}{40}
\expandafter\ifx\csname natexlab\endcsname\relax\def\natexlab#1{#1}\fi
\expandafter\ifx\csname bibnamefont\endcsname\relax
  \def\bibnamefont#1{#1}\fi
\expandafter\ifx\csname bibfnamefont\endcsname\relax
  \def\bibfnamefont#1{#1}\fi
\expandafter\ifx\csname citenamefont\endcsname\relax
  \def\citenamefont#1{#1}\fi
\expandafter\ifx\csname url\endcsname\relax
  \def\url#1{\texttt{#1}}\fi
\expandafter\ifx\csname urlprefix\endcsname\relax\def\urlprefix{URL }\fi
\providecommand{\bibinfo}[2]{#2}
\providecommand{\eprint}[2][]{\url{#2}}

\bibitem[{\citenamefont{Pines and Nozi\`{e}res}(1966)}]{Pnozieres}
\bibinfo{author}{\bibfnamefont{D.}~\bibnamefont{Pines}} \bibnamefont{and}
  \bibinfo{author}{\bibfnamefont{P.}~\bibnamefont{Nozi\`{e}res}},
  \emph{\bibinfo{title}{The Theory of Quantum Liquids}}
  (\bibinfo{publisher}{W.~A.~Benjamin, Inc}, \bibinfo{address}{New York},
  \bibinfo{year}{1966}).

\bibitem[{\citenamefont{Quong and Eguiluz}(1993)}]{quong93}
\bibinfo{author}{\bibfnamefont{A.~A.} \bibnamefont{Quong}} \bibnamefont{and}
  \bibinfo{author}{\bibfnamefont{A.~G.} \bibnamefont{Eguiluz}},
  \bibinfo{journal}{Phys. Rev. Lett.} \textbf{\bibinfo{volume}{70}},
  \bibinfo{pages}{3955} (\bibinfo{year}{1993}).

\bibitem[{\citenamefont{Aryasetiawan and Karlsson}(1994)}]{aryasetiawan94}
\bibinfo{author}{\bibfnamefont{F.}~\bibnamefont{Aryasetiawan}}
  \bibnamefont{and} \bibinfo{author}{\bibfnamefont{K.}~\bibnamefont{Karlsson}},
  \bibinfo{journal}{Phys. Rev. Lett.} \textbf{\bibinfo{volume}{73}},
  \bibinfo{pages}{1679} (\bibinfo{year}{1994}).

\bibitem[{\citenamefont{Maddocks et~al.}(1994)\citenamefont{Maddocks, Godby,
  and Needs}}]{maddocks94}
\bibinfo{author}{\bibfnamefont{N.~E.} \bibnamefont{Maddocks}},
  \bibinfo{author}{\bibfnamefont{R.~W.} \bibnamefont{Godby}}, \bibnamefont{and}
  \bibinfo{author}{\bibfnamefont{R.~J.} \bibnamefont{Needs}},
  \bibinfo{journal}{Europhys. Lett.} \textbf{\bibinfo{volume}{27}},
  \bibinfo{pages}{681} (\bibinfo{year}{1994}).

\bibitem[{\citenamefont{Fleszar et~al.}(1995)\citenamefont{Fleszar, Quong, and
  Eguiluz}}]{fleszar95}
\bibinfo{author}{\bibfnamefont{A.}~\bibnamefont{Fleszar}},
  \bibinfo{author}{\bibfnamefont{A.~A.} \bibnamefont{Quong}}, \bibnamefont{and}
  \bibinfo{author}{\bibfnamefont{A.~G.} \bibnamefont{Eguiluz}},
  \bibinfo{journal}{Phys. Rev. Lett.} \textbf{\bibinfo{volume}{74}},
  \bibinfo{pages}{590} (\bibinfo{year}{1995}).

\bibitem[{\citenamefont{Fleszar et~al.}(1997)\citenamefont{Fleszar, Stumpf, and
  Eguiluz}}]{fleszar97}
\bibinfo{author}{\bibfnamefont{A.}~\bibnamefont{Fleszar}},
  \bibinfo{author}{\bibfnamefont{R.}~\bibnamefont{Stumpf}}, \bibnamefont{and}
  \bibinfo{author}{\bibfnamefont{A.~G.} \bibnamefont{Eguiluz}},
  \bibinfo{journal}{Phys. Rev. B} \textbf{\bibinfo{volume}{55}},
  \bibinfo{pages}{2068} (\bibinfo{year}{1997}).

\bibitem[{\citenamefont{Rohlfing and Louie}(2000)}]{rohlfing00}
\bibinfo{author}{\bibfnamefont{M.}~\bibnamefont{Rohlfing}} \bibnamefont{and}
  \bibinfo{author}{\bibfnamefont{S.~G.} \bibnamefont{Louie}},
  \bibinfo{journal}{Phys.\ Rev. B} \textbf{\bibinfo{volume}{62}},
  \bibinfo{pages}{4927} (\bibinfo{year}{2000}).

\bibitem[{\citenamefont{Campillo et~al.}(1999)\citenamefont{Campillo, Rubio,
  and Pitarke}}]{campillo99}
\bibinfo{author}{\bibfnamefont{I.}~\bibnamefont{Campillo}},
  \bibinfo{author}{\bibfnamefont{A.}~\bibnamefont{Rubio}}, \bibnamefont{and}
  \bibinfo{author}{\bibfnamefont{J.~M.} \bibnamefont{Pitarke}},
  \bibinfo{journal}{Phys.\ Rev. B} \textbf{\bibinfo{volume}{59}},
  \bibinfo{pages}{12188} (\bibinfo{year}{1999}).

\bibitem[{\citenamefont{Cazalilla et~al.}(2000)\citenamefont{Cazalilla, Dolado,
  Rubio, and Echenique}}]{cazalilla00}
\bibinfo{author}{\bibfnamefont{M.~A.} \bibnamefont{Cazalilla}},
  \bibinfo{author}{\bibfnamefont{J.~S.} \bibnamefont{Dolado}},
  \bibinfo{author}{\bibfnamefont{A.}~\bibnamefont{Rubio}}, \bibnamefont{and}
  \bibinfo{author}{\bibfnamefont{P.~M.} \bibnamefont{Echenique}},
  \bibinfo{journal}{Phys.\ Rev.\ B.} \textbf{\bibinfo{volume}{61}},
  \bibinfo{pages}{8033} (\bibinfo{year}{2000}).

\bibitem[{\citenamefont{Gurtubay et~al.}(2001)\citenamefont{Gurtubay, Pitarke,
  Campillo, and Rubio}}]{gurtubay01}
\bibinfo{author}{\bibfnamefont{I.~G.} \bibnamefont{Gurtubay}},
  \bibinfo{author}{\bibfnamefont{J.~M.} \bibnamefont{Pitarke}},
  \bibinfo{author}{\bibfnamefont{I.}~\bibnamefont{Campillo}}, \bibnamefont{and}
  \bibinfo{author}{\bibfnamefont{A.}~\bibnamefont{Rubio}},
  \bibinfo{journal}{Comp. Mater. Sci.} \textbf{\bibinfo{volume}{22}},
  \bibinfo{pages}{123} (\bibinfo{year}{2001}).

\bibitem[{\citenamefont{Sch\"{o}ne and Ekardt}(2002)}]{schone02}
\bibinfo{author}{\bibfnamefont{W.~D.} \bibnamefont{Sch\"{o}ne}}
  \bibnamefont{and} \bibinfo{author}{\bibfnamefont{W.}~\bibnamefont{Ekardt}},
  \bibinfo{journal}{J.~Phys.: Condens.~Matter} \textbf{\bibinfo{volume}{14}},
  \bibinfo{pages}{4669} (\bibinfo{year}{2002}).

\bibitem[{\citenamefont{Gurtubay
  et~al.}(2004{\natexlab{a}})\citenamefont{Gurtubay, Ku, Pitarke, and
  Eguiluz}}]{gurtuhcp04}
\bibinfo{author}{\bibfnamefont{I.~G.} \bibnamefont{Gurtubay}},
  \bibinfo{author}{\bibfnamefont{W.}~\bibnamefont{Ku}},
  \bibinfo{author}{\bibfnamefont{J.~M.} \bibnamefont{Pitarke}},
  \bibnamefont{and} \bibinfo{author}{\bibfnamefont{A.~G.}
  \bibnamefont{Eguiluz}}, \bibinfo{journal}{Comp. Mater. Sci.}
  \textbf{\bibinfo{volume}{30}}, \bibinfo{pages}{104}
  (\bibinfo{year}{2004}{\natexlab{a}}).

\bibitem[{\citenamefont{Vast et~al.}(2002)\citenamefont{Vast, Reining, Olevano,
  Schattschneider, and Jouffrey}}]{tio2reining}
\bibinfo{author}{\bibfnamefont{N.}~\bibnamefont{Vast}},
  \bibinfo{author}{\bibfnamefont{L.}~\bibnamefont{Reining}},
  \bibinfo{author}{\bibfnamefont{V.}~\bibnamefont{Olevano}},
  \bibinfo{author}{\bibfnamefont{P.}~\bibnamefont{Schattschneider}},
  \bibnamefont{and} \bibinfo{author}{\bibfnamefont{B.}~\bibnamefont{Jouffrey}},
  \bibinfo{journal}{Phys.\ Rev.\ Lett.} \textbf{\bibinfo{volume}{88}},
  \bibinfo{pages}{037601} (\bibinfo{year}{2002}).

\bibitem[{\citenamefont{Gurtubay
  et~al.}(2004{\natexlab{b}})\citenamefont{Gurtubay, Ku, Pitarke, Eguiluz,
  Larson, Tischler, and Zschack}}]{gurtutio2-04}
\bibinfo{author}{\bibfnamefont{I.~G.} \bibnamefont{Gurtubay}},
  \bibinfo{author}{\bibfnamefont{W.}~\bibnamefont{Ku}},
  \bibinfo{author}{\bibfnamefont{J.~M.} \bibnamefont{Pitarke}},
  \bibinfo{author}{\bibfnamefont{A.~G.} \bibnamefont{Eguiluz}},
  \bibinfo{author}{\bibfnamefont{B.~C.} \bibnamefont{Larson}},
  \bibinfo{author}{\bibfnamefont{J.~Z.} \bibnamefont{Tischler}},
  \bibnamefont{and} \bibinfo{author}{\bibfnamefont{P.}~\bibnamefont{Zschack}},
  \bibinfo{journal}{Phys.\ Rev. B} \textbf{\bibinfo{volume}{70}},
  \bibinfo{pages}{201201(R)} (\bibinfo{year}{2004}{\natexlab{b}}).

\bibitem[{\citenamefont{Eguiluz et~al.}(2005)\citenamefont{Eguiluz, Restrepo,
  Larson, Tischler, Zshack, and Jellison}}]{nio}
\bibinfo{author}{\bibfnamefont{A.~G.} \bibnamefont{Eguiluz}},
  \bibinfo{author}{\bibfnamefont{O.~D.} \bibnamefont{Restrepo}},
  \bibinfo{author}{\bibfnamefont{B.~C.} \bibnamefont{Larson}},
  \bibinfo{author}{\bibfnamefont{J.~Z.} \bibnamefont{Tischler}},
  \bibinfo{author}{\bibfnamefont{P.}~\bibnamefont{Zshack}}, \bibnamefont{and}
  \bibinfo{author}{\bibfnamefont{G.~E.} \bibnamefont{Jellison}}
  (\bibinfo{year}{2005}), \bibinfo{note}{submitted to J. Phys. Chem. Solids}.

\bibitem[{\citenamefont{Ku et~al.}(2002)\citenamefont{Ku, Pickett, Scalettar,
  and Eguiluz}}]{ku02}
\bibinfo{author}{\bibfnamefont{W.}~\bibnamefont{Ku}},
  \bibinfo{author}{\bibfnamefont{W.~E.} \bibnamefont{Pickett}},
  \bibinfo{author}{\bibfnamefont{R.~T.} \bibnamefont{Scalettar}},
  \bibnamefont{and} \bibinfo{author}{\bibfnamefont{A.~G.}
  \bibnamefont{Eguiluz}}, \bibinfo{journal}{Phys. Rev. Lett.}
  \textbf{\bibinfo{volume}{88}}, \bibinfo{pages}{057001}
  (\bibinfo{year}{2002}).

\bibitem[{\citenamefont{Grenier et~al.}(2005)\citenamefont{Grenier, Hill,
  Kiryukhin, Ku, Kim, Thomas, Cheong, Tokura, Tomioka, Casa
  et~al.}}]{grenier05}
\bibinfo{author}{\bibfnamefont{S.}~\bibnamefont{Grenier}},
  \bibinfo{author}{\bibfnamefont{J.~P.} \bibnamefont{Hill}},
  \bibinfo{author}{\bibfnamefont{V.}~\bibnamefont{Kiryukhin}},
  \bibinfo{author}{\bibfnamefont{W.}~\bibnamefont{Ku}},
  \bibinfo{author}{\bibfnamefont{Y.-J.} \bibnamefont{Kim}},
  \bibinfo{author}{\bibfnamefont{K.~J.} \bibnamefont{Thomas}},
  \bibinfo{author}{\bibfnamefont{S.-W.} \bibnamefont{Cheong}},
  \bibinfo{author}{\bibfnamefont{Y.}~\bibnamefont{Tokura}},
  \bibinfo{author}{\bibfnamefont{Y.}~\bibnamefont{Tomioka}},
  \bibinfo{author}{\bibfnamefont{D.}~\bibnamefont{Casa}}, \bibnamefont{et~al.},
  \bibinfo{journal}{Phys. Rev. Lett.} \textbf{\bibinfo{volume}{94}},
  \bibinfo{pages}{047203} (\bibinfo{year}{2005}).

\bibitem[{\citenamefont{Adler}(1962)}]{adler62}
\bibinfo{author}{\bibfnamefont{S.~L.} \bibnamefont{Adler}},
  \bibinfo{journal}{Phys. Rev.} \textbf{\bibinfo{volume}{126}},
  \bibinfo{pages}{413} (\bibinfo{year}{1962}).

\bibitem[{\citenamefont{Raether}(1980)}]{raether80}
\bibinfo{author}{\bibfnamefont{H.}~\bibnamefont{Raether}},
  \emph{\bibinfo{title}{Excitation of Plasmons and Interband Transitions by
  Electrons}}, vol.~\bibinfo{volume}{88} of \emph{\bibinfo{series}{Springer
  Tracts in Modern Physics}} (\bibinfo{publisher}{Springer-Verlag},
  \bibinfo{address}{Berlin}, \bibinfo{year}{1980}).

\bibitem[{\citenamefont{Misell and Atkins}(1973)}]{misell73}
\bibinfo{author}{\bibfnamefont{D.~L.} \bibnamefont{Misell}} \bibnamefont{and}
  \bibinfo{author}{\bibfnamefont{A.~J.} \bibnamefont{Atkins}},
  \bibinfo{journal}{Philos. Mag. 27} \textbf{\bibinfo{volume}{27}},
  \bibinfo{pages}{95} (\bibinfo{year}{1973}).

\bibitem[{\citenamefont{Eisenberger et~al.}(1973)\citenamefont{Eisenberger,
  Platzman, and Pandy}}]{platzman73}
\bibinfo{author}{\bibfnamefont{P.}~\bibnamefont{Eisenberger}},
  \bibinfo{author}{\bibfnamefont{P.~M.} \bibnamefont{Platzman}},
  \bibnamefont{and} \bibinfo{author}{\bibfnamefont{C.}~\bibnamefont{Pandy}},
  \bibinfo{journal}{Phys. Rev. Lett.} \textbf{\bibinfo{volume}{31}},
  \bibinfo{pages}{311} (\bibinfo{year}{1973}).

\bibitem[{\citenamefont{Platzman et~al.}(1992)\citenamefont{Platzman, Isaacs,
  Williams, Zschack, and Ice}}]{platzman92}
\bibinfo{author}{\bibfnamefont{P.~M.} \bibnamefont{Platzman}},
  \bibinfo{author}{\bibfnamefont{E.~D.} \bibnamefont{Isaacs}},
  \bibinfo{author}{\bibfnamefont{H.}~\bibnamefont{Williams}},
  \bibinfo{author}{\bibfnamefont{P.}~\bibnamefont{Zschack}}, \bibnamefont{and}
  \bibinfo{author}{\bibfnamefont{G.~E.} \bibnamefont{Ice}},
  \bibinfo{journal}{Phys. Rev. B} \textbf{\bibinfo{volume}{46}},
  \bibinfo{pages}{12943} (\bibinfo{year}{1992}).

\bibitem[{\citenamefont{Sch\"{u}lke et~al.}(1995)\citenamefont{Sch\"{u}lke,
  Schmitz, Schulte-Screpping, and Kaprolat}}]{schulke95}
\bibinfo{author}{\bibfnamefont{W.}~\bibnamefont{Sch\"{u}lke}},
  \bibinfo{author}{\bibfnamefont{J.~R.} \bibnamefont{Schmitz}},
  \bibinfo{author}{\bibfnamefont{H.}~\bibnamefont{Schulte-Screpping}},
  \bibnamefont{and} \bibinfo{author}{\bibfnamefont{A.}~\bibnamefont{Kaprolat}},
  \bibinfo{journal}{Phys. Rev. B} \textbf{\bibinfo{volume}{52}},
  \bibinfo{pages}{11721} (\bibinfo{year}{1995}).

\bibitem[{\citenamefont{Larson et~al.}(1996)\citenamefont{Larson, Tischler,
  Isaacs, Zschack, Fleszar, and Eguiluz}}]{larson96}
\bibinfo{author}{\bibfnamefont{B.~C.} \bibnamefont{Larson}},
  \bibinfo{author}{\bibfnamefont{J.~Z.} \bibnamefont{Tischler}},
  \bibinfo{author}{\bibfnamefont{E.~D.} \bibnamefont{Isaacs}},
  \bibinfo{author}{\bibfnamefont{P.}~\bibnamefont{Zschack}},
  \bibinfo{author}{\bibfnamefont{A.}~\bibnamefont{Fleszar}}, \bibnamefont{and}
  \bibinfo{author}{\bibfnamefont{A.~G.} \bibnamefont{Eguiluz}},
  \bibinfo{journal}{Phys. Rev. Lett.} \textbf{\bibinfo{volume}{77}},
  \bibinfo{pages}{1346} (\bibinfo{year}{1996}).

\bibitem[{\citenamefont{Macrander et~al.}(1996)\citenamefont{Macrander,
  Montano, Price, Kushnir, Blasdell, Kao, and Cooper}}]{macrander96}
\bibinfo{author}{\bibfnamefont{A.~T.} \bibnamefont{Macrander}},
  \bibinfo{author}{\bibfnamefont{P.~A.} \bibnamefont{Montano}},
  \bibinfo{author}{\bibfnamefont{D.~L.} \bibnamefont{Price}},
  \bibinfo{author}{\bibfnamefont{V.~I.} \bibnamefont{Kushnir}},
  \bibinfo{author}{\bibfnamefont{R.~C.} \bibnamefont{Blasdell}},
  \bibinfo{author}{\bibfnamefont{C.~C.} \bibnamefont{Kao}}, \bibnamefont{and}
  \bibinfo{author}{\bibfnamefont{B.~R.} \bibnamefont{Cooper}},
  \bibinfo{journal}{Phys. Rev. B} \textbf{\bibinfo{volume}{54}},
  \bibinfo{pages}{305} (\bibinfo{year}{1996}).

\bibitem[{\citenamefont{Montano et~al.}(2002)\citenamefont{Montano, Price,
  Macrander, and Cooper}}]{montano02}
\bibinfo{author}{\bibfnamefont{P.~A.} \bibnamefont{Montano}},
  \bibinfo{author}{\bibfnamefont{D.~L.} \bibnamefont{Price}},
  \bibinfo{author}{\bibfnamefont{A.~T.} \bibnamefont{Macrander}},
  \bibnamefont{and} \bibinfo{author}{\bibfnamefont{B.~R.}
  \bibnamefont{Cooper}}, \bibinfo{journal}{Phys. Rev. B}
  \textbf{\bibinfo{volume}{66}}, \bibinfo{pages}{165218}
  (\bibinfo{year}{2002}).

\bibitem[{\citenamefont{Kohn and Sham}(1965)}]{KS65}
\bibinfo{author}{\bibfnamefont{W.}~\bibnamefont{Kohn}} \bibnamefont{and}
  \bibinfo{author}{\bibfnamefont{L.}~\bibnamefont{Sham}},
  \bibinfo{journal}{Phys. Rev.} \textbf{\bibinfo{volume}{140}},
  \bibinfo{pages}{A11333} (\bibinfo{year}{1965}).

\bibitem[{\citenamefont{Dreizler and Gross}(1990)}]{dreizler90}
\bibinfo{author}{\bibfnamefont{R.~M.} \bibnamefont{Dreizler}} \bibnamefont{and}
  \bibinfo{author}{\bibfnamefont{E.~K.~U.} \bibnamefont{Gross}},
  \emph{\bibinfo{title}{Density Functional Theory}}
  (\bibinfo{publisher}{Springer-Verlag}, \bibinfo{address}{Berlin},
  \bibinfo{year}{1990}).

\bibitem[{\citenamefont{Singh}(1994)}]{singh94}
\bibinfo{author}{\bibfnamefont{D.}~\bibnamefont{Singh}},
  \emph{\bibinfo{title}{Pseudopotentials and the LAPW Method}}
  (\bibinfo{publisher}{Kluwer Academic Publishers}, \bibinfo{address}{Boston,
  Dordrecht, London}, \bibinfo{year}{1994}).

\bibitem[{\citenamefont{Blaha et~al.}(1999)\citenamefont{Blaha, Schwarz, and
  Luitz}}]{wien97}
\bibinfo{author}{\bibfnamefont{P.}~\bibnamefont{Blaha}},
  \bibinfo{author}{\bibfnamefont{K.}~\bibnamefont{Schwarz}}, \bibnamefont{and}
  \bibinfo{author}{\bibfnamefont{J.}~\bibnamefont{Luitz}},
  \emph{\bibinfo{title}{WIEN97, A Full Potential Linearized Augmented Plane
  Wave Package for Calculating Crystal Properties}},
  \bibinfo{address}{Technische Universit\"{a}t Wien, Austria}
  (\bibinfo{year}{1999}), \bibinfo{note}{{I}SBN 3-9501031-0-4}.

\bibitem[{\citenamefont{Ku}(2000)}]{weiphd}
\bibinfo{author}{\bibfnamefont{W.}~\bibnamefont{Ku}}, Ph.D. thesis,
  \bibinfo{school}{The University of Tennessee}, \bibinfo{address}{Knoxville}
  (\bibinfo{year}{2000}).

\bibitem[{\citenamefont{Runge and Gross}(1984)}]{runge-gross84}
\bibinfo{author}{\bibfnamefont{E.}~\bibnamefont{Runge}} \bibnamefont{and}
  \bibinfo{author}{\bibfnamefont{E.~K.~U.} \bibnamefont{Gross}},
  \bibinfo{journal}{Phys. Rev. Lett.} \textbf{\bibinfo{volume}{52}},
  \bibinfo{pages}{997} (\bibinfo{year}{1984}).

\bibitem[{\citenamefont{Gross and Kohn}(1985)}]{gross-kohn85}
\bibinfo{author}{\bibfnamefont{E.~K.~U.} \bibnamefont{Gross}} \bibnamefont{and}
  \bibinfo{author}{\bibfnamefont{W.}~\bibnamefont{Kohn}},
  \bibinfo{journal}{Phys. Rev. Lett.} \textbf{\bibinfo{volume}{55}},
  \bibinfo{pages}{2850} (\bibinfo{year}{1985}).

\bibitem[{\citenamefont{Petersilka et~al.}(1996)\citenamefont{Petersilka,
  Gossmann, and Gross}}]{petersilka96}
\bibinfo{author}{\bibfnamefont{M.}~\bibnamefont{Petersilka}},
  \bibinfo{author}{\bibfnamefont{U.~J.} \bibnamefont{Gossmann}},
  \bibnamefont{and} \bibinfo{author}{\bibfnamefont{E.~K.~U.}
  \bibnamefont{Gross}}, \bibinfo{journal}{Phys. Rev. Lett.}
  \textbf{\bibinfo{volume}{76}}, \bibinfo{pages}{1212} (\bibinfo{year}{1996}).

\bibitem[{\citenamefont{Callen and Welton}(1951)}]{callen51}
\bibinfo{author}{\bibfnamefont{H.~B.} \bibnamefont{Callen}} \bibnamefont{and}
  \bibinfo{author}{\bibfnamefont{T.~R.} \bibnamefont{Welton}},
  \bibinfo{journal}{Phys. Rev.} \textbf{\bibinfo{volume}{83}},
  \bibinfo{pages}{34} (\bibinfo{year}{1951}).

\bibitem[{\citenamefont{Tischler et~al.}(2003)\citenamefont{Tischler, Larson,
  Zschack, Fleszar, and Eguiluz}}]{alnorm}
\bibinfo{author}{\bibfnamefont{J.~Z.} \bibnamefont{Tischler}},
  \bibinfo{author}{\bibfnamefont{B.~C.} \bibnamefont{Larson}},
  \bibinfo{author}{\bibfnamefont{P.}~\bibnamefont{Zschack}},
  \bibinfo{author}{\bibfnamefont{A.}~\bibnamefont{Fleszar}}, \bibnamefont{and}
  \bibinfo{author}{\bibfnamefont{A.~G.} \bibnamefont{Eguiluz}},
  \bibinfo{journal}{~Phys.~Stat.~Sol. (b)} \textbf{\bibinfo{volume}{237}},
  \bibinfo{pages}{280} (\bibinfo{year}{2003}).

\bibitem[{\citenamefont{Perdew and Wang}(1992)}]{Pwang}
\bibinfo{author}{\bibfnamefont{J.~P.} \bibnamefont{Perdew}} \bibnamefont{and}
  \bibinfo{author}{\bibfnamefont{Y.}~\bibnamefont{Wang}},
  \bibinfo{journal}{Phys. Rev. B} \textbf{\bibinfo{volume}{45}},
  \bibinfo{pages}{13244} (\bibinfo{year}{1992}).

\bibitem[{\citenamefont{Ceperley and Alder}(1980)}]{CA80}
\bibinfo{author}{\bibfnamefont{D.~M.} \bibnamefont{Ceperley}} \bibnamefont{and}
  \bibinfo{author}{\bibfnamefont{B.~J.} \bibnamefont{Alder}},
  \bibinfo{journal}{Phys. Rev. Lett.} \textbf{\bibinfo{volume}{45}},
  \bibinfo{pages}{566} (\bibinfo{year}{1980}).

\bibitem[{\citenamefont{Brousseau-Lahaye
  et~al.}(1975)\citenamefont{Brousseau-Lahaye, Colliex, Frandon, Gasgnier, and
  Trebbia}}]{sceels75}
\bibinfo{author}{\bibfnamefont{B.}~\bibnamefont{Brousseau-Lahaye}},
  \bibinfo{author}{\bibfnamefont{C.}~\bibnamefont{Colliex}},
  \bibinfo{author}{\bibfnamefont{J.}~\bibnamefont{Frandon}},
  \bibinfo{author}{\bibfnamefont{M.}~\bibnamefont{Gasgnier}}, \bibnamefont{and}
  \bibinfo{author}{\bibfnamefont{P.}~\bibnamefont{Trebbia}},
  \bibinfo{journal}{~Phys.~Stat.~Sol. (b)} \textbf{\bibinfo{volume}{69}},
  \bibinfo{pages}{257} (\bibinfo{year}{1975}).

\bibitem[{\citenamefont{Montano and Macrander}(2000)}]{montano00}
\bibinfo{author}{\bibfnamefont{P.~A.} \bibnamefont{Montano}} \bibnamefont{and}
  \bibinfo{author}{\bibfnamefont{A.~T.} \bibnamefont{Macrander}},
  \bibinfo{journal}{J.~Phys.~Chem.~Solids} \textbf{\bibinfo{volume}{61}},
  \bibinfo{pages}{415} (\bibinfo{year}{2000}).

\end{thebibliography}

\end{document}